\documentclass[onecolumn,draftclsnofoot]{IEEEtran}

\usepackage[table]{xcolor}
\usepackage{cite}
\makeatletter
\renewcommand\OverciteFont{\fontsize\ssf@size\baselineskip\selectfont}
\makeatother
\usepackage{wrapfig}
\usepackage[cmex10]{amsmath}
\usepackage{amssymb}
\usepackage{amsthm}
\usepackage{booktabs}
\usepackage{bm}
\usepackage{algorithm,algpseudocode}
\usepackage{multirow}
\usepackage{tikz}
\usetikzlibrary{math,arrows,matrix,positioning,calc}
\usepackage{pgfplots}
\usepackage{enumitem}
\usepackage{subfigure}

\usepackage[hidelinks]{hyperref}
 \newtheorem{theorem}{Theorem}

 \newtheorem{lemma}[theorem]{Lemma}
 
\theoremstyle{definition}
 \newtheorem{definition}{Definition}

\theoremstyle{remark}
\newtheorem{remark}{Remark}

\newcommand{\ff}{\mathbb{F}}
\newcommand{\diff}{\mathrm{d}}
\newcommand{\cv}[1]{\mathbf{#1}}
\newcommand{\mc}[1]{\mathcal{#1}}
\newcommand{\E}{\mathbf{E}}
\newcommand{\rank}{\mathrm{rank}}
\newcommand{\ee}{\mathrm{Er}}

\newcommand{\Qin}{\mc Q_{\mathrm{i}}}
\newcommand{\Qout}{\mc Q_{\mathrm{o}}}
\newcommand{\Sin}{\mc S_{\mathrm{i}}}
\newcommand{\Sout}{\mc S_{\mathrm{o}}}
\DeclareMathOperator*{\argmax}{arg\,max} 

\definecolor{longhorn}{rgb}{0.8, 0.33, 0.0}

\begin{document}

\title{On Achievable Rates of Line Networks with Generalized Batched
  Network Coding}
\author{Jie~Wang,~\IEEEmembership{Graduate Student Member,}~Shenghao~Yang,~\IEEEmembership{Member,~IEEE,}~Yanyan~Dong~and~Yiheng~Zhang
  \thanks{This work was supported in part by the
National Key R\&D Program of China (No. 2022YFA1005000) and by National Natural Science Foundation of China (No. 62171399 and No. 12141108). (Corresponding author: Shenghao Yang.)}
  \thanks{J.~Wang is with the H. Milton Stewart School of Industrial
    and Systems Engineering, Georgia Institute of Technology, Atlanta 30332,
    USA (jwang3163@gatech.edu).}%
  \thanks{S.~Yang is with the School of Science and Engineering, The Chinese
    University of Hong Kong, Shenzhen, Shenzhen 518172, China
    (shyang@cuhk.edu.cn).}%
  \thanks{Y.~Dong is with  the School of Science and Engineering, The Chinese
    University of Hong Kong, Shenzhen, Shenzhen 518172, China (yanyandong@link.cuhk.edu.cn).} %
  \thanks{Y.~Zhang is with the Information Networking Institute, Carnegie
    Mellon University, Pittsburgh 15289, USA (yihengzh@andrew.cmu.edu).}
}

\maketitle

\begin{abstract}
To better understand the wireless network design with a large number of hops, we investigate a line network formed by general discrete memoryless channels (DMCs), which may not be identical. Our focus lies on Generalized Batched Network Coding (GBNC) that encompasses most existing schemes as special cases and achieves the min-cut upper bounds as the parameters batch size and inner block length tend to infinity.
The inner blocklength of GBNC provides upper bounds on the required latency and buffer size at intermediate network nodes. By employing a ``bottleneck status'' technique, we derive new upper bounds on the achievable rates of GBNCs These bounds surpass the min-cut bound for large network lengths when the inner blocklength and batch size are small.
For line networks of canonical channels, certain upper bounds hold even with relaxed inner blocklength constraints.
Additionally, we employ a ``channel reduction'' technique to generalize the existing achievability results for line networks with identical DMCs to networks with non-identical DMCs.
For line networks with packet erasure channels, we make refinement in both the upper bound and the coding scheme, and showcase their proximity through numerical evaluations.
\end{abstract}


\section{Introduction}

We investigate multi-hop line topology networks formed by concatenating discrete memoryless channels (DMCs), which are fundamental channel models in communication systems. In this line network, the first node serves as the source node, the last node serves as the destination node, and the intermediate nodes establish connections between them. Multi-hop wireless communication networks find applications in diverse domains, including underwater acoustic networks~\cite{zhang2010analysis}, free space optical communication~\cite{tang2014multihop}, deep space communication networks~\cite{huakai2016simplified}, field area networks~\cite{harada2017ieee}, and terahertz communications~\cite{bhardwaj2022performance}.

In the absence of constraints on storage and latency at the intermediate nodes, the network capacity is determined by the min-cut from the source to the destination, achievable through the hop-by-hop implementation of capacity-achieving channel codes~\cite{cover06}. However, as the number of hops increases, the hop-by-hop coding approach introduces significant communication latency and storage requirements at the intermediate nodes, which are critical factors in multi-hop wireless networks~\cite{bedewy2019age, farazi2019fundamental}.
In their work~\cite{Niesen2007}, Niesen, Fragouli, and Tuninetti investigated the line network capacity by considering a fixed inner blocklength $N$ at the intermediate nodes. This blocklength has an impact on delay and buffer size. Assuming identical channels in the line network (referred to as $Q$), and when the zero-error capacity of $Q$ is non-zero, they demonstrated that using a constant $N$ allows achieving any constant rate below the zero-error capacity for any given number of hops $L$. Conversely, when the zero-error capacity of $Q$ is zero, a class of codes with a constant $N$ can achieve rates on the order of $\Omega(e^{-c L})$, where $c$ is a constant. Additionally, if $N$ is of the order of $\ln L$, it is possible to achieve any rate below the capacity of $Q$.

However, despite these achievability results, the min-cut remains the strongest upper bound for line networks. It is still uncertain whether the diminishing achievable rates observed with increasing network length are fundamental or if there exist more efficient coding strategies that can achieve higher rates. Furthermore, it is worth exploring the possibility of reducing the processing latency and buffer size requirements beyond the complexity of $O(N)$. With these inquiries in mind, we embark on a comprehensive investigation of line networks formed by DMCs.

Improving the general upper bound for multi-hop networks is an extremely challenging task, as suggested in the network information theory literature~\cite{gamma11}. In this paper, our focus is on a specific class of codes called Generalized Batched Network Coding (GBNC).
While batched network coding has been extensively studied for networks of packet erasure channels~\cite{Silva2009, Heidarzadeh2010, yaoli11, yang14bats, bin_expander15, bin18ldpc}, we extend batched network coding to accommodate general DMCs, which may not be identical. GBNC, introduced in \S\ref{sec:batched} of this paper, consists of an outer code and an inner code. The outer code encodes information messages into batches of coded symbols, while the inner code performs recoding operations within each batch. GBNC incorporates two key parameters: the batch size $M$ and the inner blocklength $N$.
There are several reasons that make GBNC well-suited for our research objectives. Firstly, GBNC encompasses a wide range of codes as special cases. The coding scheme examined in~\cite{Niesen2007} corresponds to GBNC with $M=N$. Both decode-and-forward and retransmission schemes can be viewed as special inner codes for GBNC. Secondly, when both $M$ and $N$ can be arbitrarily large, GBNC has the capability to achieve the min-cut. Lastly, GBNC enables us to explicitly characterize latency and buffer size. Our formulation reveals that the recoding latency and buffer size at an intermediate node are upper-bounded by a linear order of $N$.

In this paper, we derive both upper and lower bounds on the achievable
rate of GBNC in terms of the parameters $M$, $N$, and network length
$L$. Compared to our previous conference
papers~\cite{yang2019capacity, yang2020upper}, the main results
presented in this paper are either improved or entirely new.  Using a
``bottleneck status'' technique, we obtain new upper bounds on the
achievable rate of GBNC for line networks consisting of channels with
$0$ zero-error capacity.  We begin by proving the converses for a
class of channels known as \emph{canonical channels}, which are
characterized by having an output symbol that occurs with a positive
probability for all possible input symbols, and then extend the
results to non-canonical channels (detailed in \S\ref{sec:conv}) We
demonstrate through various cases that our upper bounds outperform the
min-cut.

To gain a more explicit understanding, we conduct further analysis on
how the upper and lower bounds scale with $L$ for different scenarios of $M$ and
$N$. Notably, when $N=O(1)$, our upper bound reveals that the
achievable rate must decay exponentially with $L$, aligning with the
achievable rates obtained in~\cite{Niesen2007}. By utilizing a
``channel reduction'' technique (detailed in \S\ref{sec:relatedscheme}
and \S\ref{sec:reduction}), we extend the achievability results
of~\cite{Niesen2007} to line networks with non-identical DMCs.
Additionally, when $N=O(\ln L)$ and $M=O(1)$, our upper bound
indicates that the achievable rate is $O(1/\ln L)$, which is a new
scalability compared with the previous ones obtained
in~\cite{Niesen2007}.  We demonstrate that rates of $\Omega(1/\ln L)$
can be attained using $M=O(1)$ and $N=O(\ln L)$. In a general
decode-and-forward approach, a buffer size of $O(\ln L)$ is
required. However, specific codes enable a reduced buffer size of
$O(\ln\ln L)$ (refer to \S\ref{sec:rep}). To exemplify this result, we
consider a repetition coding scheme, which prompts us to explore
simpler schemes for line networks with a large number of hops. A
summarization of the scalability results can be found in
Table~\ref{tab:1}.

In the context of line networks with packet erasure channels, we make advancements in both the upper bound and the coding scheme. Through extensive numerical evaluations, we establish a close proximity between the upper bound and the achievable rates of the coding scheme (see \S\ref{sec:era}). This finding serves as motivation for future research endeavors aimed at improving the upper bound and developing more efficient coding schemes tailored to specific channel characteristics.

Last, our results are extended to networks where certain channels have a positive zero-error capacity (see  \S\ref{sec:ex}).

\begin{table}[tb]
  \caption{Summarization of the achievable rate scalability for the channels with $0$ zero-error capacity using batched codes. Here, $c$ and $c'$ have constant values that do not change with $L$. The upper/lower bound marked with $*$ is obtained in this paper.}
  \label{tab:1}
    \centering
  \subtable[upper bound]{
    \rowcolors{3}{gray!30}{white}
    \begin{tabular}{ccccc}
      \toprule
      batch size $M$ & inner blk-length $N$  & buffer size & upper bound \\
      \midrule
      unbounded & $O(1)$ & unbounded  &  {$O(e^{-c'L})$}$^{*}$ \\
      $O(1)$ & $\Omega(\ln L)$ & unbounded & {$O(1/\ln L)$}$^{*}$ \\
      unbounded & unbounded  & unbounded & {$O(1)$}$^{\text{\cite{cover06}}}$\\
      \bottomrule
    \end{tabular}
  }

  \subtable[lower bound]{
    \rowcolors{3}{gray!30}{white}
    \begin{tabular}{cccc}
      \toprule
      batch size $M$ & inner blk-length $N$ & buffer size & lower bound \\
      \midrule
      $O(1)$ & $O(1)$ &  $O(1)$ & {$\Omega(e^{-cL})$}$^{\text{\cite{Niesen2007}},*}$ \\
      $O(1)$ & $O(\ln L)$ & $O(\ln\ln L)$ & {$\Omega(1/\ln L)$}$^{*}$ \\
      $O(\ln L)$ & $O(\ln L)$ & $O(\ln L)$ & {$\Omega(1)$}$^{\text{\cite{Niesen2007}},*}$  \\
      \bottomrule
    \end{tabular}
   }
\end{table}

Throughout this paper, we use $\log$ to denote the logarithm of base $2$, and $\ln$ to denote the natural logarithm of base $e$.
For random variables represented by uppercase letters (e.g., $X$), we use the corresponding lowercase letters (e.g., $x$) to represent their instances.
We use $P$ to denote the probability of events, and we may write $P(X=x)$ as $P(x)$ to simplify the notation. We use $p_X$ to denote the probability mass function of the discrete random variable $X$, where subscripts may be omitted.
Most of the notations used throughout this manuscript are given in Table~\ref{tab:notation:I} for easy of reference.
All omitted proofs can be found in the supplementary material online~\cite{wang23jsacsupp}.

\begin{table}[t]
  \caption{Some notations used in the paper, listed in the alphabetical order.}\label{tab:notation:I}%
  \centering %
  \begin{tabular}{lp{6.5cm}}
    \toprule
    Notation & Explanation \\ %
    \midrule %
   $\mathcal{A}$ &  Batch alphabet. \\
   $C(Q)$            &  Channel capacity of channel $Q$. \\ %
   $C_0(Q)$        &  Zero-error capacity of channel $Q$. \\ %
   $C_L(M,N)$     & Maximum achievable rate of all recoding schemes with batch size $M$ and inner blocklength $N$. \\ %
   $E_{0,\ell}$      & Event that all $N$ outputs of $Q_{\ell}$ are equal to the same value regardless of channel input. \\ %
   $E_0$				& Event that there exists one link $\ell$ such that $E_{0,\ell}$ holds. \\ %
  $ \ee_{\ell}$ &        Coding error exponent for channel $Q_{\ell}$. \\ %
  $\ee^*$		& Smallest coding error exponent among all $\ell\ge1$.\\ %
    $L$                   &  Network length. \\ %
   $M$                   &  Batch size.\\
   $N$                   &   Inner blocklength.\\
   $Q_\ell$					&   Discrete memoryless channel of link $\ell$.\\
   $\cv U_{\ell}/\cv Y_{\ell}$  & The input/output of $N$ uses of the $\ell$-th communication link.\\
   $W_L$				& End-to-end transition matrix of the batch channel from $\cv X$ to $\cv Y_L$.\\
  $\cv X\in \mathcal{A}^M$  & A generic batch.\\
  $\cv X{[k]}$   		&   The $k$-th entry in $\cv X$.\\
  $\cv Z_{\ell}$		&  Channel status of $Q_{\ell}$.\\
    \bottomrule
  \end{tabular}%
\end{table}

\section{Line Networks and Generalized Batched Network Coding}
\label{sec:batched}
In this section, we describe the line network model and introduce batched network coding.

\subsection{Line Network Model}
\label{sec:line}

A line network of length $L$ consists of nodes labeled as $0, 1, \ldots, L$, with directed communication links from node ${\ell-1}$ to node $\ell$. Each link is a discrete memoryless channel (DMC) with fixed finite input and output alphabets $\Qin$ and $\Qout$ respectively. The transition matrix for link $\ell$ is denoted as $Q_\ell$. The line network is formed by concatenating $Q_1, Q_2, \ldots, Q_L$.
This study focuses on communication between the first node, referred to as the \emph{source node}, and the last node, known as the \emph{destination node}. The nodes numbered $1, 2, \ldots, L-1$ are referred to as the \emph{intermediate nodes}.

Let $C(Q)$ and $C_0(Q)$ denote the channel capacity and the zero-error capacity of a DMC with transition matrix $Q$ respectively. Without any constraints at the network nodes, the capacity of the network is given by $\min_{\ell=1}^L C(Q_{\ell})$, which is also known as the \emph{min-cut}. Achieving the min-cut involves using a capacity achieving code at each hop, where intermediate nodes decode the previous link's code and encode the message using the next link's code. This scheme is commonly referred to as \emph{decode-and-forward}.
However, as we will discuss later, decode-and-forward is not always the optimal solution when considering both latency and buffer size at the intermediate nodes.
Next, we present a general coding scheme for the line network and examine the relationship between the coding parameters and latency as well as buffer size.

\subsection{Generalized Batched Network Coding}
\label{sec:bat}

A \emph{Generalized Batched Network Code} (GBNC) comprises an outer code and an inner code. The outer code, executed at the source node, encodes a message from a finite set and generates multiple batches, each containing $M$ symbols from a finite set $\mathcal{A}$. The parameter $M$ is known as the \emph{batch size}.
The inner code operates on individual batches separately, employing recoding operations at nodes $0, 1, \ldots, {L-1}$.

Let's define the recoding process for a generic batch $\cv X \in \mc A^M$. At the source node, the recoding transforms the original $M$ symbols of $\cv X$ into $N$ recoded symbols $\cv U_1$ in $\Qin$, where $N$ is a positive integer referred to as the \emph{inner blocklength}. The recoding at the source node is represented by the function $\phi_0: \mc A^M \rightarrow \Qin^N$, such that $\cv U_1 = \phi_0(\cv X)$.

At an intermediate node $\ell$, recoding is performed on the $N$ received symbols $\cv Y_{\ell} \in \Qout^N$ to generate $N$ recoded symbols $U_{\ell+1}\in \Qin^N$ for transmission on the outgoing link of node $\ell$. Due to the memoryless property of $Q_{\ell}$, the conditional probability of $\cv Y_{\ell}=\cv y$ given $\cv U_{\ell}=\cv u$ is
\begin{equation}\label{eq:ch}
P(\cv Y_{\ell}=\cv y|\cv U_{\ell}=\cv u) = Q_\ell^{\otimes N}(\cv y | \cv u) \triangleq \prod_{i=1}^N Q_\ell(\cv y[i]|\cv u[i]),
\end{equation}
where $\cv y[k]$ ($1\leq k\leq N$) represents the $k$th entry in $\cv y$.
The recoding at node $\ell$ is represented by the function $\phi_{\ell}: \Qout^N \rightarrow \Qin^N$, such that $\cv U_{\ell+1} = \phi_{\ell}(\cv Y_{\ell})$.
In general, the number of recoded symbols transmitted by different nodes can vary~\cite{Wang2021, yin2021unified}. However, for simplicity, we assume they are all the same for the analysis.

At the destination node, all received symbols, which may belong to different batches, are jointly decoded. The inner code's end-to-end operation, with the given recoding function $\phi_{\ell}$ at all nodes, can be viewed as a memoryless channel referred to as a \emph{batch channel}, which takes $\cv X$ as the input and produces $\cv Y_L$ as the output. Fig.~\ref{fig:rec} illustrates the variables involved in the recoding process, forming the Markov chain:
\begin{equation}
\label{eq:1}
\cv X \rightarrow \cv U_1 \rightarrow \cv Y_1 \rightarrow \cdots \rightarrow \cv U_L \rightarrow \cv Y_L.
\end{equation}
The end-to-end transition matrix $W_L$ of the batch channel can be derived using $\phi_{\ell}$ and $Q_{\ell}$.

\begin{figure}[tb]
  \centering
  \begin{tikzpicture}[font=\footnotesize, x=2cm,
    dot/.style={circle,draw=gray!80,fill=gray!20,thick,inner
      sep=1pt,minimum size=12pt}]
    \node[dot,label=below:{node $0$}] (s) {$\cv X$};
    \node[dot,label=below:{node $1$},right=0.6 of s] (a) {};
    \draw[->] (s)-- node[pos=0.2,above] {$\cv U_1$} node[pos=0.8,above] {$\cv Y_1$} node[pos=0.5,below] {$Q_1$} (a);
    \node[right=0.3 of a] (b) {} edge[<-] node[pos=0.8,above] {$\cv U_2$} node[pos=0.2,below] {$Q_2$} (a);
    \node[right=-0.1 of b] (e) {$\cdots$};
    \node[dot,label=below:{node $\ell$},right=0.3 of e] (f) {} edge[<-] node[above] {$\cv Y_{\ell}$} (e);
    \node[right=0.3 of f] (g) {} edge[<-] node[above,pos=0.8] {$\cv U_{\ell+1}$} (f);
    \node[right=-0.1 of g] (h) {$\cdots$};
    \node[dot,label=below:{node $L-1$},right=0.3 of h] (i) {} edge[<-] node[pos=0.2,above] {$\cv Y_{L-1}$} (h);
    \node[dot,label=below:{node $L$},right=0.6 of i] (t) {};
    \draw[->] (i)-- node[pos=0.2,above] {$\cv U_{L}$} node[pos=0.8,above] {$\cv Y_{L}$} node[pos=0.5,below] {$Q_L$} (t);
  \end{tikzpicture}
  \caption{A line network with the random variables involved in recoding.}\label{fig:rec}
\end{figure}

The outer code serves as a channel code for the batch channel $W_L$ to ensure end-to-end reliability. Given a recoding scheme $\{\phi_{\ell}\}$, the maximum achievable rate of the outer code is $\max_{p_{\cv X}} I(\cv X;\cv Y_{L})$ for $N$ channel uses, where $p_{\cv X}$ represents the distribution of $\cv X$. The objective of designing a recoding scheme, given parameters $M$ and $N$, is to maximize $\frac{1}{N}\max_{p_{\cv X}} I(\cv X;\cv Y_{L})$. Let $C_L(M,N)$ denote the maximum achievable rate among all recoding schemes with batch size $M$ and inner blocklength $N$, defined as:
\begin{equation}\label{eq:recoding}
C_L(M,N) = \max_{\{\phi_{\ell}\},p_{\cv X}} \frac{I(\cv X;\cv Y_{L})}{N} =  \max_{\{\phi_{\ell}\}, p_{\cv X}} \frac{I(p_{\cv X}, W_L)}{N}.
\end{equation}
$C_L(M,N)$ is also referred to as the capacity of GBNCs with parameters $M$ and $N$. We can then maximize $C_L(M,N)$ while considering constraints on $M$ and $N$, which impact both the recoding latency and the buffer size.

Recoding functions $\{\phi_\ell\}$ can generally be random. However, the convexity of $I(p_{\cv X}, W_L)$ for a fixed $p_{\cv X}$ with respect to $W_L$ implies the existence of a deterministic recoding scheme that achieves $C_L(M,N)$. In particular, the coding scheme analyzed in \cite{Niesen2007} considers the case where $M=N$. A special inner code known as decode-and-forward will be discussed in \S\ref{sec:ach}.
GBNCs generalize the batched network codes studied for networks with packet erasure channels in literature (see discussion in \S\ref{sec:era}).

\subsection{Buffer Size and Latency at Intermediate Nodes}

Let's now delve into the buffer size requirement and latency at the intermediate nodes in GBNCs. In this discussion, we consider a sequential transmission model where symbols of a batch are transmitted consecutively.  We will discuss the buffer size required for caching the received symbols for recoding at an intermediate node, as well as the latency between receiving the first symbol of a batch and transmitting the first symbol of the same batch. We will disregard the space and time costs associated with executing recoding $\phi_\ell$.

The key principle of GBNCs is the independent application of recoding to each batch. In the worst case scenario, an intermediate node begins transmitting the first recoded symbol of a batch only after receiving all $N$ symbols of that batch. Consequently, the latency of a batch at an intermediate node is upper bounded by $O(N)$. Since an intermediate node can only transmit symbols of a batch after receiving at least one symbol from that batch, the lower bound on the latency at an intermediate node is $1$. The accumulated end-to-end recoding latency across all intermediate nodes falls within the range of $\Omega(L)$ to $O(NL)$.

Similarly, in the worst-case scenario, an intermediate node starts transmitting the first recoded symbol of a batch only after receiving all $N$ symbols of that batch. Additionally, these received symbols need to be cached for $N$ more channel uses. Therefore, an intermediate node needs to cache at most $2N$ symbols: $N$ symbols of the batch for transmitting and $N$ symbols of the same batch for receiving. This indicates that the buffer size required for caching symbols at an intermediate node is $O(N)$.

\section{Converse for Line Networks of Channels with \texorpdfstring{$0$}{0} Zero-error Capacity}
\label{sec:conv}

One known upper bound of $C_L(M,N)$ is the min-cut $\min_{\ell=1}^L C(Q_\ell)$. However, this bound may not be sufficient for small values of $M$ and $N$.
When $C_0(Q_{\ell}) = 0$ for all $\ell$, 
in this section, we introduce a technique called a ``bottleneck status'' to derive a potentially tighter bound on $C_L(M,N)$ when $M$ and $N$ are small.

The bottleneck status refers to an event $E_0$ that is associated with the channel $W_L$ and is independent of $\cv X$. Let
\begin{subequations}
  \label{eq:wl01}
    \begin{align}
W_L^{(0)}(\cv y\mid\cv x) &=  P(\cv Y_L=\cv y\mid\cv X=\cv x, E_0),\label{eq:wl01:a}\\
W_L^{(1)}(\cv y\mid\cv x) &= P(\cv Y_L=\cv y\mid\cv X=\cv x, \overline{E_0}).\label{eq:wl01:b}
\end{align}
\end{subequations}
The channel $W_L$ can be expressed as $  W_L=W_L^{(0)}p_0 + W_L^{(1)}p_1$, where $p_0=P(E_0), p_1=P(\overline{E_0})$.
As mutual information $ I(p_{\cv X}, W_L) $ is convex w.r.t. $W_L$ for given $p_{\cv X}$, we can establish the upper bound as follows:
\begin{equation} \label{eq:bot}
I(p_{\cv X}, W_L) \leq p_0I(p_{\cv X}, W_L^{(0)})  + p_1I(p_{\cv X}, W_L^{(1)}).
\end{equation}
The crucial step is to design the event $E_0$ in order to obtain the desired upper bound.

\begin{definition}
For $0<\varepsilon\leq 1$, we call a DMC $Q:\Qin \to \Qout$ an \emph{$\varepsilon$-canonical channel} if there exists $y^* \in \Qout$ such that for every $x \in \Qin$, $Q(y^*|x)\geq \varepsilon$.  
\end{definition}
For a canonical channel, there exists an output symbol $y^*$ that occurs with a positive probability for all the inputs. The binary erasure channel (BEC) and binary symmetric channel (BSC) are both canonical channels, but a typewriter channel is non-canonical. Note that a canonical $Q$ has $C_0(Q)=0$. 
We first introduce our technique to design a bottleneck status for canonical channels, and then discuss the general channels.

\subsection{Line Network of Canonical Channels}
\label{sec:canonical}

In this subsection, we study a line network consisting of $\varepsilon$-canonical channels $Q_{\ell}, \ell=1,\ldots, L$.
To design the bottleneck status $E_0$, we adopt a formulation of DMCs in \cite[\S7.1]{yeung08}. %
Define $Z = (Z[x], x\in \Qin)$, where $Z[x], x\in \Qin$ are independent random variables on $\Qout$ with the distribution
$P(Z[x] = y) = Q(y|x)$.
The relation between the input $X$ and output $Y$ of a DMC $Q$ can be modeled as 
\begin{equation}\label{eq:ssd} 
  Y = \alpha(X, Z=(Z[x], x\in \Qin)) \triangleq  \sum_{x\in\Qin}\bm 1\{X=x\}Z[x],
\end{equation}
where $\bm 1$ denotes the indicator function. Here
$Z = (Z[x], x\in \Qin)$ is also called \emph{channel status variable}, and $\alpha$ is called the \emph{channel function}.
We denote by $\alpha_{\ell}$ the channel function of $Q_{\ell}$.

Consider a GBNC with inner blocklength $N$ for the line network. 
With the alternative channel formulation~\eqref{eq:ssd}, we can write for $\ell=1,\ldots,L$, and $i=1,\ldots, N$,
 $ \cv Y_{\ell}[i] = \alpha_{\ell}(\cv U_{\ell}[i], \cv Z_{\ell}[i]).$
Here $\cv Z_{\ell}[i]= (\cv Z_{\ell}[i,x], x\in \Qin)$ is the channel status variable for the $i$th use of the channel $Q_{\ell}$, where
\begin{equation}
  \label{eq:cs1}
  P(\cv Z_{\ell}[i,x] = y) = Q_{\ell}(y|x).
\end{equation}
Define $\cv Z_{\ell}=(\cv Z_{\ell}[i], i=1,\ldots,N)$. 
For notation simplicity, we rewrite the channel relation as
\begin{equation}\label{eq:chz}
  \cv Y_{\ell} = \alpha_{\ell}^{(N)}(\cv U_{\ell}, \cv Z_{\ell}).
\end{equation}

Given that $Q_{\ell}$ is $\varepsilon$-canonical, there exists an output denoted as $y_{\ell}^*$ satisfying
\begin{equation}
  \label{eq:2}
  Q_{\ell}(y_{\ell}^*|x)\geq \varepsilon \text{ for all } x\in \Qin.
\end{equation}
Let's define
\begin{equation}
  \label{eq:5}
  E_{0,\ell} =\{\cv Z_{\ell}[i,x] = y_{\ell}^*, i\in \{1,\ldots, N\}, x\in \Qin\}.
\end{equation}
Under the condition $E_{0,\ell}$, all $N$ outputs of $Q_{\ell}$ are equal to $y_{\ell}^*$ for any possible channel input, rendering the channel useless. We can quantify the probability of $E_{0,\ell}$ as follows:
\begin{IEEEeqnarray}{rCl}
P(E_{0,\ell})&=&\prod_{\substack{i\in\{1,\ldots,N\}, x\in\Qin}} P(\cv Z_{\ell}[i,x]=y_{\ell}^*) \label{eq:8sd} \\
&=& \prod_{\substack{i\in\{1,\ldots,N\}, x\in\Qin}} Q_{\ell}(y_{\ell}^*| x) \\
& \ge & \varepsilon^{|\Qin|N}, \label{eq:e0l}
\end{IEEEeqnarray}
where \eqref{eq:8sd} follows from \eqref{eq:cs1}, and \eqref{eq:e0l} follows from \eqref{eq:2}.
Now we define the bottleneck status
\begin{equation}
  \label{eq:e0}
  E_0 = \lor_{\ell=1}^LE_{0,\ell}.
\end{equation}
This event implies the existence of at least one link $\ell$ in the network that is deemed useless and hence the network is useless.

\begin{lemma}\label{lemma:3}
  When $Q_{\ell}$, $\ell = 1,\ldots, L$ are all $\varepsilon$-canonical channels,
  for $W_L^{(0)}$ defined in \eqref{eq:wl01:a} and $E_0$ defined in \eqref{eq:e0}, $I(p_{\cv X}, W_L^{(0)}) = 0$. 
\end{lemma}

\begin{lemma}\label{lemma:w1}
  When $Q_{\ell}$, $\ell = 1,\ldots, L$ are all $\varepsilon$-canonical channels,
  for $W_L^{(1)}$ defined in \eqref{eq:wl01:b} and $E_0$ defined in \eqref{eq:e0}, we have 1)
  \begin{equation}
    \label{eq:3}
    P(\overline{E_0}) \leq (1-\varepsilon^{|\Qin| N})^L,
  \end{equation}
  2) for any $\ell=1,\ldots,L$
  \begin{equation}
    \label{eq:7}
    I(p_{\cv X}, W_L^{(1)}) \leq  \max_{p_{\cv U_\ell}} I(\cv U_\ell;\cv Y_{\ell} \mid \overline{E_{0,\ell}}).
  \end{equation}
\end{lemma}

In Lemma~\ref{lemma:w1}, $\max_{p_{\cv U_{\ell}}} I(\cv U_\ell;\cv Y_{\ell} \mid \overline{E_{0,\ell}})$ is the capacity of the channel $Q_{\ell}^N$ under the condition $\overline{E_{0,\ell}}$.
One upper bound is $\frac{1}{N}\max_{p_{\cv U_\ell}} I(\cv U_\ell;\cv Y_{\ell} \mid \overline{E_{0,\ell}}) \leq \log \min(|\Qin|,|\Qout|)$.
In the following lemma, we give a better upper bound that converges $C(Q_{\ell})$ when $N$ tends to infinity.

\begin{lemma}\label{lemma:upb}
  Consider a channel $Q$ as defined in \eqref{eq:ssd} by $(\alpha, Z)$. Fix an output $y^*$ such that $Q(y^*|x) = P(Z[x]=y^*) \geq \epsilon$ for all input $x$, where $\epsilon>0$. For $N$ uses of the channel, let $Z[i,x]$ be the channel variable of the $i$th uses associated with the input $x$.  Let $E_0$ be the event that $\{Z[i,x] = y^*, i=1,\ldots,N, x\in \Qin\}$. Let $W$ be the channel formed by $N$ uses of $Q$ under the condition of $\overline{E_0}$. 
  Let 
  \begin{equation}
    \label{eq:8}
    D(Q,N) =    (q^*+p_0) \log \frac{q^*-p_0}{\epsilon^N-p_0} + q^* \log\frac{ \epsilon^N}{q^*}
  \end{equation}
  where
$p_0= \left(\prod_{x} Q(y^*|x)\right)^N$, and $q^* = \max_{\cv x} Q^{\otimes N}(\cv y^*|\cv x)$. Then
  \begin{equation}
    \label{eq:6}
    \frac{1}{N} I(p,W)\le C^*(Q,N) \triangleq \frac{1}{1-p_0} \left( C(Q) + \frac{D(Q,N)}{N} \right).
  \end{equation}
\end{lemma}

Based on the relation~\eqref{eq:bot},
together with Lemmas~\ref{lemma:3}, \ref{lemma:w1}, and \ref{lemma:upb}, we derive the following theorem.

\begin{theorem}\label{thm:ca}
  Consider a length-$L$ line network of $\varepsilon$-canonical channels with finite input and output alphabets $\Qin$ and $\Qout$, respectively. 
The capacity of GBNCs with batch size $M$ and inner blocklength $N$ has the following upper bound:
  \begin{equation}
    \label{eq:1ca}
    \begin{aligned}
C_L(M,N)\leq (1-&\varepsilon^{|\Qin| N})^L \min \Big\{ C^*(Q_\ell,N),\\
&\log|\Qin|, \log|\Qout|,
    \frac{M\log|\mc A|}{N}\Big\}.
    \end{aligned}
  \end{equation}
Moreover,
  \begin{enumerate}
\item when $N=O(1)$, $\max C_L(M,N) = O((1-\varepsilon^{|\Qin| N})^L)$;
\item when $M=O(1)$, $\max C_L(M,N) = O(1/\ln L)$;
\item when $M$ and $N$ are arbitrary, $\max C_L(M,N) = O(1)$.
\end{enumerate}
\end{theorem}
\begin{IEEEproof}%
Recall the capacity of GBNC in \eqref{eq:recoding}, where 
  \begin{IEEEeqnarray}{rCl}
    I(p_{\cv X}, W_L) & \leq & p_0I(p_{\cv X}, W_L^{(0)})  + p_1I(p_{\cv X}, W_L^{(1)})  \label{eq:41} \\
    & = &  p_1I(p_{\cv X}, W_L^{(1)}) \label{eq:42} \\
    & \leq & (1-\varepsilon^{|\Qin| N})^L  I(p_{\cv X}, W_L^{(1)}),  \label{eq:43} 
  \end{IEEEeqnarray}
  where \eqref{eq:41} follows from \eqref{eq:bot}, \eqref{eq:42} is obtained by applying Lemma~\ref{lemma:3}, and \eqref{eq:43} follows from Lemma~\ref{lemma:w1}-1).
  The upper bound in \eqref{eq:1ca} is proved by 
  \begin{IEEEeqnarray}{rCl}
    I(p_{\cv X}, W_L^{(1)}) & \leq & H(\cv X) \leq M \log |\mc A| \\
    I(p_{\cv X}, W_L^{(1)}) & \leq & I(\cv U_\ell;\cv Y_{\ell} \mid \overline{E_{0,\ell}}) \leq N \log \min(|\Qin|, |\Qout|) \quad \quad \label{eq:d8s} \\
      I(p_{\cv X}, W_L^{(1)})
    & \leq &  N C^*(Q_\ell,N), \label{eq:44}
  \end{IEEEeqnarray}
  where \eqref{eq:d8s} follows from Lemma~\ref{lemma:w1}-2) and \eqref{eq:44} holds due to Lemma~\ref{lemma:upb}.

  The remainder part of the theorem is proved by analyzing the upper bound in \eqref{eq:1ca} for different values of $M$ and $N$. In particular, Case~2) is obtained using the following Lemma~\ref{lemma:1}.
\end{IEEEproof} 

\begin{lemma}\label{lemma:1}
  For fixed real number $0<\epsilon<1$ and integer $L> 1$,
  the function $F(N) = (1-\epsilon^N)^L/N$ of integer $N$ is maximized when $N$ is $\Theta(\ln L)$, and  the optimal value of
  $F(N)$ is
  $\Theta\left(\frac{\ln (1/\epsilon)}{\ln L}\right)$.
\end{lemma}

\begin{figure}
  \centering
  \subfigure[the capacity upper bound~\eqref{eq:1ca} with inner blocklength $N\in\{1,2,3\}$ and batch size $M=3$]{
    \includegraphics[width=0.4\textwidth]{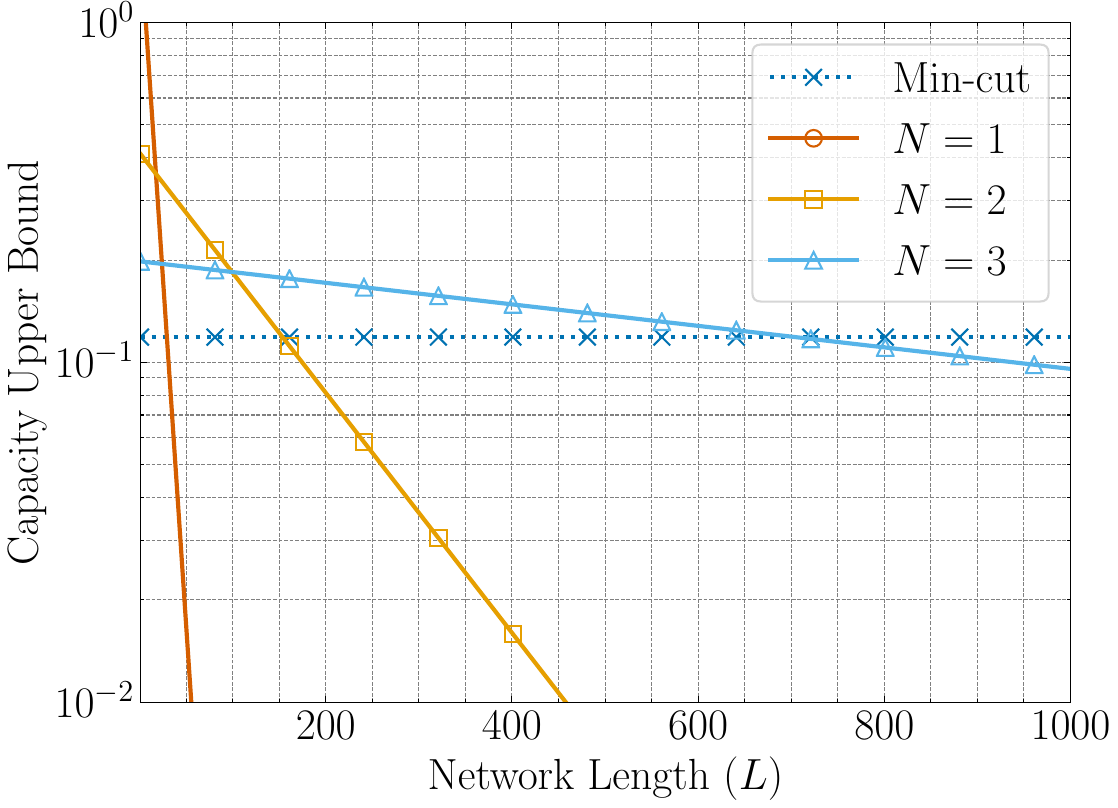}}
  \subfigure[the capacity upper bound~\eqref{eq:1ca} with inner blocklength $N=\lfloor10*\log(L+1)\rfloor$ and batch size $M\in\{1,2,3\}$]{
\includegraphics[width=0.4\textwidth]{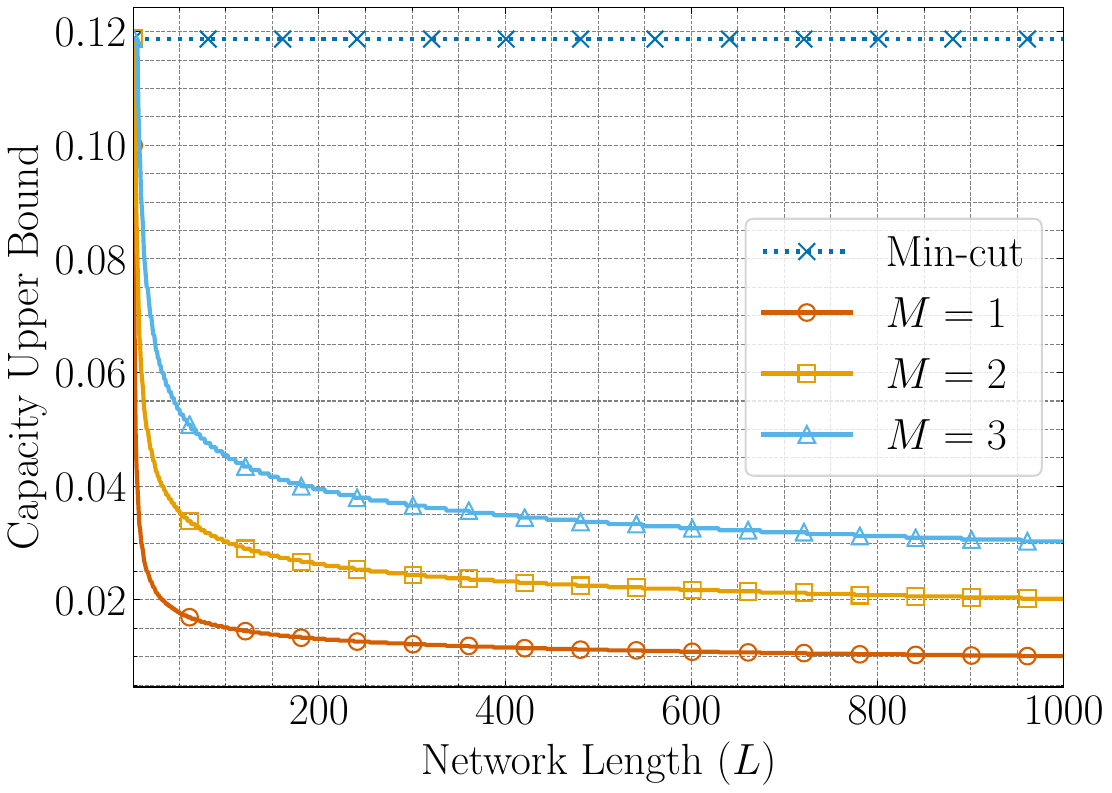}}
\caption{
Numerical illustrations of the capacity upper bound of GBNC using BSC with crossover probability $\epsilon=0.3$ and $|\mc A|=2$.}
        \label{fig:numerical:BSC:upper}
\end{figure}

  To illustrate the capacity upper bound in Theorem~\ref{thm:ca}, we evaluate it for the network formed by BSCs in Fig.~\ref{fig:numerical:BSC:upper}, and use the min-cut for baseline comparison.
Fig.~\ref{fig:numerical:BSC:upper}(a) depicts, for each hop length $L$, the upper bound~\eqref{eq:1ca} when $M,N=O(1)$.
It reveals the exponential decay of the capacity with respect to $L$, and the min-cut is in geneal a loose upper bound for sufficiently large $L$.
Fig.~\ref{fig:numerical:BSC:upper}(b) shows the upper bound~\eqref{eq:1ca} when $M=O(1), N=O(\ln L)$. 
In this case, the capacity decays slowly as $L$ increases, and the min-cut is a loose upper bound as well.

\subsection{General Channels}
\label{sec:genc}

Consider a channel $Q:\Qin \to \Qout$ with $C_0(Q)=0$, modeled as in  \eqref{eq:ssd}.
Since $Q$ may not be canonical, there may not exist an output symbol that occurs with a positive probability for all inputs.
Furthermore, if $Q$ is non-canonical, $Q^{\otimes m}$ is also non-canonical for any positive integer $m$. For instance, let's define the channel $Q_{3\times 3}$ with $\Qin=\Qout = \{0,1,2\}$ and
$Q_{3\times 3}(0|0) = Q_{3\times 3}(0|1) = Q_{3\times 3}(1|0) = Q_{3\times 3}(1|2) = Q_{3\times 3}(2|1) = Q_{3\times 3}(2|2) = 1/2$. We can check that $Q_{3\times 3}^{\otimes m}$ is non-canonical.
Consequently, the bottleneck status we observe for a canonical channel cannot be directly extended to non-canonical channels.

\begin{figure}%
\centering
\begin{tikzpicture}[font=\footnotesize, scale=0.8,
  dot/.style={draw,fill,circle,inner sep=0pt,minimum size=4pt}]
    \foreach \y in {0,1,2}
    {
      \foreach \x in {0,1,2,3}
      {
        \node[dot] (a\x\y) at (2*\x,-\y) {};
      }
      \node[left=2pt of a0\y] {$\y$};
      \draw[->] (a1\y) -- (a2\y);
    }
    \foreach \x/\z in {0/1,2/3}
    {  
      \draw[->,dashed] (a\x0) -- (a\z0);
      \draw[->,dashed] (a\x0) -- (a\z1);
      \draw[->,dashed] (a\x1) -- (a\z0);
      \draw[->,dashed] (a\x1) -- (a\z2);
      \draw[->,dashed] (a\x2) -- (a\z2);
      \draw[->,dashed] (a\x2) -- (a\z1);
      \node at ($(a\x2)!0.5!(a\z2)+(0,-0.4)$) {$Q_{3\times 3}$};
    }
    \node at ($(a12)!0.5!(a22)+(0,-0.4)$) {$\Phi$};
  \end{tikzpicture}
 \caption{Concatenation of two $Q_{3\times 3}$'s with recoding $\Phi$. The end-to-end channel is given by $W=Q_{3\times 3}\Phi Q_{3\times 3}$. Here $\Phi$ is a deterministic transition matrix with $\Phi(i|i)=1$. The transition from an input to an output connected by a dashed has  probability $1/2$. Any output of $W$ can occur with a positive probability for all inputs.}
  \label{fig:c3}
\end{figure}

To investigate the converse of general channels, we employ a technique that involves concatenating multiple channels through recoding, resulting in a new channel that is canonical. Let's use the example of $Q_{3\times 3}$ to illustrate this idea. We consider the concatenation of two copies of $Q_{3\times 3}$ using a $3\times 3$ \emph{deterministic} transition matrix $\Phi$, yielding the new channel $W = Q_{3\times 3} \Phi Q_{3\times 3}.$
In this setup, $\Phi$ maps an output of the first channel as an input of the second channel.
Refer to the illustration in Fig.~\ref{fig:c3}. For the first channel, it is guaranteed that at one of the output in the set $\{0,1\}$ occurs with a positive probability for any input. Recoding $\Phi$ can map the outputs $0$ and $1$ of the first channel to either the same input or two distinct inputs of the second channel. Due to the properties of $Q_{3\times 3}$, regardless of the specific mapping, there will always exist an output of $W$ that occurs with a positive probability for any input of $W$.

Now we discuss the general case. 
For a channel $Q:~\Qin\to\Qout$, denote by $\varepsilon_Q$ the maximum value such that for any $x, x'\in \Qin$, there exists $y\in \Qout$ such that $Q(y|x)\geq \varepsilon_{Q}$ and $Q(y|x')\geq \varepsilon_{Q}$. 
In the case of $Q_{3\times 3}$, we have $\varepsilon_{Q_{3\times 3}} = 1/2$. 
Note that $\varepsilon_Q>0$ if and only if $C_0(Q)=0$ (see~\cite{shannon1956}).
Since $C_0(Q_{\ell})=0$, it is possible to observe the same output for any two channel inputs of $Q_{\ell}$. Exploiting this property, we can prove that for any subset $\Sin$ of $\Qin$, there exists a subset $\Sout$ of $\Qout$ with a size less than half of $\Sin$, such that for any input in $\Sin$, it is possible to observe an output in $\Sout$. This can be formally stated as the following lemma.

\begin{lemma}\label{lemma:z}
  Consider a DMC $Q:\Qin \to \Qout$ with $\varepsilon_Q>0$ modelled by $(\alpha, Z)$. For any non-empty set $\Sin\subseteq\Qin$, there exist a subset $\mc Z$ of the range of $Z$ and a subset $\Sout\subseteq\Qout$ with $|\Sout|\le \lceil |\Sin|/2 \rceil$ such that $\alpha(x, z)\in \Sout$ for any $x\in \Sin$ and $z\in \mc Z$, and $P(Z \in \mc Z) \geq \varepsilon_{Q}^{|\Sin|}$.
\end{lemma}

Based on the aforementioned lemma, we can concatenate a sufficiently large number of  consecutive channels in a line network to create a canonical channel. In order to establish the upper bound, we need to demonstrate that for a certain $\epsilon > 0$ and any recoding schemes, a number of consecutive channels in the line network form an $\epsilon$-canonical channel. The following lemma provides justification for this feasibility.

\begin{lemma}\label{lemma:concatenate}
  Let $  K=\lceil N \log|\Qin| \rceil.$
  Consider a line network of $K$ DMCs $Q_{\ell}$ with $\varepsilon_{Q_{\ell}}\geq \epsilon >0$.
  For any deterministic GBNC with the inner blocklength $N$ and the recoding functions $\{\phi_{\ell}\}$,
  let $G = Q_{1}^{\otimes N}\phi_{1} Q_{2}^{\otimes N} \cdots \phi_{K-1}Q_{K}^{\otimes N}.$ Then $G$ is $\varepsilon^{N(2|\Qin|^N+K)}$-canonical.
\end{lemma}
\begin{IEEEproof}%
Consider a deterministic GBNC as described in \S\ref{sec:batched}. Channel $Q_{\ell}^{\otimes N}$ can be modelled by the function $\alpha_{\ell}^N$ with the channel status variable $Z_{\ell}$ as in \eqref{eq:chz}. 
As $\varepsilon_{Q_{\ell}^{\otimes N}} \geq \varepsilon_{Q_{\ell}}^{N} \geq \epsilon^N> 0$, the condition of applying Lemma~\ref{lemma:z} on $Q_{\ell}^{\otimes N}$ is satisfied.

Let $\Sin^{(1)} = \Qin^N$. Applying Lemma~\ref{lemma:z} on $Q_{1}^{\otimes N}$ w.r.t. $\Sin^{(1)}$, there exists subsets $\mc Z^{(1)}$ of the range of $Z_1$ and $\Sout^{(1)}\subseteq \Qout^N$ with $|\Sout^{(1)}| \leq \lceil |\Sin^{(1)}|/2 \rceil$ such that $\alpha_{1}^{N}(\cv x, z_1) \in \Sout^{(1)}$ for any $\cv x \in \Sin^{(1)}$ and $z_1\in \mc Z^{(1)}$, and $P(Z_1\in \mc Z^{(1)}) \geq \varepsilon^{N|\Qin|^N}$.

For $i= 2,3,\ldots,K$, define recursively
\begin{equation}
  \Sin^{(i)} = \left\{ \phi_{i-1}(\cv y): \cv y \in \Sout^{(i-1)}\right\},
\end{equation}
and $\Sout^{(i)}$ and $\mc Z^{(i)}$ as in the proof of Lemma~\ref{lemma:z} w.r.t. $Q_{i}^{\otimes N}$ and $\Sin^{(i)}$ so that $\alpha_{i}^{\otimes N}(\cv x,z) \in \Sout^{(i)}$ for any $\cv x \in \Sin^{(i)}$ and $z\in \mc Z^{(i)}$, and $P(Z_i\in \mc Z^{(i)}) \geq \varepsilon^{N|\Sin^{(i)} |}$.
According to the construction, $|\Sin^{(i)}| \leq |\Sout^{(i-1)}|$ and 
$|\Sout^{(i)}| \leq \lceil |\Sin^{(i)}|/2 \rceil$.
Hence
$|\Sout^{(K)}| \leq \lceil |\Sin^{(1)}|/2^{K} \rceil = 1$. 
Since the set $\Sout^{(K)}$ is non-empty, 
we have $|\Sout^{(K)}|=1$, i.e., there exists an output of $Q_{K}^{\otimes N}$ that occurs with a positive probability for all inputs of $Q_{1}^{\otimes N}$.

Under the condition $Z_i\in \mc Z^{(i)}, i=1,\ldots,K$, the output of $G$ must be unique for all possible channel inputs. Note that
\begin{equation}\label{eq:pxs}
  P(Z_i\in \mc Z^{(i)}, i=1,\ldots,K) \geq \varepsilon^{N\sum_{i=1}^K|\Sin^{(i)}|} \geq \varepsilon^{N(2|\Qin|^N+K)}.
\end{equation}
The proof is completed.
\end{IEEEproof}

Based on the aforementioned lemma, we are now ready to prove the upper bound for the general case. The main idea is to divide the line network into consecutive segments, each consisting of $K$ consecutive channels. Lemma~\ref{lemma:concatenate} guarantees that each segment can form a canonical channel.
In contrast to the proof of Theorem~\ref{thm:ca}, the key difference lies in the definition of the bottleneck status. In this case, we can utilize $Z_i\in \mc Z^{(i)}$ in the proof of Lemma~\ref{lemma:concatenate} to define the bottleneck status. This demonstrates another way of applying the bottleneck status technique.

\begin{theorem}\label{thm:scalability}
Consider a length-$L$ line network of channels $\{Q_{\ell}\}_{\ell=1}^L$ with finite input and output alphabets and  $\varepsilon_{Q_{\ell}}\geq \varepsilon>0$ for all $\ell$.  When $L> N \log |\Qin|$,  
the capacity of GBNCs with batch size $M$ and inner blocklength $N$ has the following upper bound:
\begin{equation}
    \label{eq:main}
    \begin{aligned}
C_L(M,N)\leq& (1-\varepsilon^{N(2|\Qin|^N+K)})^{\lfloor L/K\rfloor }\\
&\cdot\min\{ M/N \log |\mc A|, \log |\Qin|, \log |\Qout| \},
    \end{aligned}
\end{equation}
where $K=\lceil N \log|\Qin| \rceil$.
Moreover, 
\begin{enumerate}
\item when $N=O(1)$, $\max C_L(M,N) = O((1-\varepsilon')^L)$ for certain $\varepsilon'\in (0,1)$;
\item when $M=O(1)$ and $N=\Omega(\ln L)$, $\max C_L(M,N) =  O(1/\ln L)$;
\item when $M$ and $N$ are arbitrary, $\max C_L(M,N) = O(1)$.
\end{enumerate}
\end{theorem}
\begin{IEEEproof}%
  Let $L' = \lfloor L/K \rfloor$. As $L> N\log|\Qin|$, we have
  $L'\geq 1$.  Consider a GBNC as described in \S\ref{sec:batched}.
  Without loss of optimality, we assume a deterministic recoding
  scheme, i.e., $\phi_{\ell}$ are deterministic.
  For $i=2,\ldots,L'$,  define
  \begin{equation*}
     G_i = Q_{K(i-1)+1}^{\otimes N}\phi_{K(i-1)+1} Q_{K(i-1)+2}^{\otimes N} \cdots \phi_{Ki-1}Q_{Ki}^{\otimes N}.
   \end{equation*}
   According to Lemma~\ref{lemma:concatenate}, we know that $G_i$, $i=2,\ldots, L'$ are all $\varepsilon^{N(2|\Qin|^N+K)}$-canonical and forms a length-$L'$ network.
Let  $  \tilde W_{L'} = \phi_0 G_1 \phi_{K} G_2 \phi_{2K}\cdots G_{L'},$
which is the end-to-end transition matrix of a GBNC with inner blocklength $1$ for the length-$L'$ network of canonical channels $G_i$. By the data processing inequality, $  I(p_{\cv X}, W_L) \leq I(p_{\cv X}, \tilde W_{L'}).$

Fix an $\ell\in \{1,2,\ldots,L'\}$. Considering the sets $\mc Z^{(i)},
   i=1,\ldots,K$ in the proof of Lemma~\ref{lemma:concatenate} for
   $G_{\ell}$, define
   \begin{equation*}
     E_{0,\ell} = \{Z_i\in \mc Z^{(i)}, i=1,\ldots,K\}.
   \end{equation*}
   Define the bottleneck status $E_{0} =
   \land_{\ell=1}^{L'} E_{0,\ell}$. Let $p_1 = P(\overline{E_{0}})$
   and $p_0=1-p_1$.  Using this bottleneck status $E_0$,
   we can define $\tilde W_{L'}^{(0)}$ and $\tilde W_{L'}^{(1)}$ as in \eqref{eq:wl01}.
   Similar as the proof of Theorem~\ref{thm:ca}, we have
   \begin{IEEEeqnarray}{rCl}
    I(p_{\cv X}, \tilde W_{L'}) & \leq & p_0I(p_{\cv X}, \tilde W_{L'}^{(0)})  + p_1I(p_{\cv X}, \tilde W_{L'}^{(1)})  \label{eq:g41} \\
    & = &  p_1I(p_{\cv X}, \tilde W_{L'}^{(1)}) \label{eq:g42} \\
    & \leq & (1-\varepsilon^{N(2|\Qin|^N+K)})^L  I(p_{\cv X},\tilde W_{L'}^{(1)})  \label{eq:g43} 
  \end{IEEEeqnarray}
  where \eqref{eq:g43} follows from \eqref{eq:pxs}. The proof is completed by bounding $I(p_{\cv X},\tilde W_{L'}^{(1)})$ using the alphabet size.
\end{IEEEproof}

\begin{remark}
Theorem~\ref{thm:ca} provides stronger results for line networks of canonical channels compared to Theorem~\ref{thm:scalability}. The upper bound given in \eqref{eq:1ca} is strictly better than the one in \eqref{eq:main}.
For general channels, it is possible to further improve Theorem~\ref{thm:scalability} by enhancing Lemma~\ref{lemma:upb}. However, when directly applying Lemma~\ref{lemma:upb} to canonical channels $G_{\ell}$ in the proof of Theorem~\ref{thm:scalability}, the resulting $D(G_{\ell},1)$ depends on the specific GBNC employed. In order to prove an upper bound that holds independently of the chosen GBNC, it would be necessary to establish a GBNC-independent upper bound on $D(G_{\ell},1).$ This matter is not discussed in the current paper.
\end{remark}

\section{Achievable Rates using Decode-and-Forward}
\label{sec:ach}

In this section, we discuss the lower bounds of the achievable rates of line networks.
We will first study the achievable rates when $N=O(\ln L)$ using two recoding schemes: decode-and-forward and repetition, which can achieve different scalability of the buffer size. 
When $N=O(1)$, for a line network of identical channels, a rate that exponentially decays with $L$ can be achieved as proved in~\cite{Niesen2007}. We will extend their results for line networks where channels may not be identical. 

\subsection{Decode-and-forward Recoding}
\label{sec:relatedscheme}

We discuss a class of GBNC recoding called
\emph{decode-and-forward}. When there is a trivial outer code,
decode-and-forward has been extensively studied and widely applied in
the existing communication systems~\cite{gamma11}.  We first describe
decode-and-forward recoding in the GBNC framework, and then
discuss the achievable rates.  

Following the notations in \S\ref{sec:bat}, we consider a GBNC with batch size $M$. Let
$(f_\ell,g_\ell)$ be a channel code for $Q_\ell$ where $f_\ell:\mc A^M\rightarrow \Qin^N$ and $g_{\ell}:\Qout^N\rightarrow \mc A^M$ are the encoding and decoding functions, respectively. Consider the transmission
of a generic batch $\cv X$. The source node transmits
$\cv U_1 = f_1(\cv X)$.  Each intermediate node $\ell$ first receives $\cv Y_{\ell}$ and then
transmits $\cv U_{\ell+1} = f_{\ell+1}(g_{\ell}(\cv Y_{\ell}))$. In
other words, the recoding function $\phi_\ell$ behaves as follows:
\begin{itemize}
\item For $i=1,\ldots,N$, the node $\ell$ just keeps the received symbols in the buffer.
Therefore, the buffer size is $\Theta(N)$. 
\item After receiving the $N$ symbols of $\cv Y_{\ell}$, the node $\ell$ generates $f_{\ell+1}(g_{\ell}(\cv Y_{\ell}))$. If the decoding is correct at nodes $1,\ldots,\ell$, then $g_{\ell}(\cv Y_{\ell}) = \cv X$ and $\cv U_{\ell+1} = f_{\ell+1}(\cv X)$.
\end{itemize}
Let $\epsilon_{\ell}$ denote the maximum decoding error probability of $(f_\ell,g_{\ell})$ for $Q_{\ell}$. Due to the fact that if the decoding is correct at all the nodes $ 1,\ldots,L $, it holds that $ g_L(\cv Y_L) = \cv X $, we have 
\begin{equation}\label{eq:22}
	P(g_L(\cv Y_L) = \cv X) \geq \prod_{\ell=1}^L(1-\epsilon_{\ell}).
\end{equation}
Let $C = \min_{\ell=1}^LC(Q_{\ell})$ be the min-cut of the line network.
When $\frac{M}{N}\log |\mc A|<C$ and $N$ is sufficiently large, by the channel coding theorem of DMCs, there exists $(f_\ell,g_{\ell})$ such that $\epsilon_{\ell}$ can be arbitrarily small. 
This gives us the well-known result that the min-cut $C$ is achievable using decode-and-forward recoding when $M$ and $N$ are allowed to be arbitrarily large~\cite{cover06}.

When all the channels are identical, it has been shown that if $M=\Theta(N)$ and $N=O(\ln L)$, a constant rate lower than $C$ can be achieved by GBNC~\cite{Niesen2007}. 
We briefly rephrase their discussion for the case where the channels of the line network are not necessarily identical. 
Consider a sequence of DMCs $Q_{\ell},\ell=1,2,\ldots$ with $C = \inf\{C(Q_{\ell}),\ell\geq 1\}>0$.
Suppose parameters $M$ and $N$ are chosen to satisfy $
\frac{M}{N}\log |\mc A|\in[0,C]$.
Using random coding arguments~\cite{gallager68}, there exists $(f_\ell,g_{\ell})$ such that
\begin{equation}\label{eq:23}
  \epsilon_{\ell} \leq \exp(-N \ee_{\ell}(r)),
\end{equation}
where $\ee_{\ell}$ is the random coding error exponent for $Q_{\ell}$.
For certain $0<C' \leq C$, assume $\ee^*(r)\triangleq \inf\{\ee_{\ell}(r),\ell\geq 1\} >0$ for all $0\leq r < C'$. 
The following theorem shows the achievable rate of decode-and-forward recoding scheme.

\begin{theorem}\label{Theorem:lower:bound:decoding:forward}
For the line network of length $L$, where the $\ell$th link is $Q_{\ell}$, the GBNC with decode-and-forward recoding scheme, batch size $M$, and inner blocklength $N$ achieves rate
\begin{equation}
  \begin{aligned}
C_L(M,N)\ge 
  \frac{M\log|\mc A|}{N}&\left(
  1 - e^{-N\ee^*(M\log|\mc A|/N)}
  \right)^L - \frac{1}{N}.
\end{aligned}
\label{Eq:lower:bound:CL}
\end{equation}
Moreover, 
 \begin{enumerate}
 	\item when $M= \Theta(N)$ and $N = O(\ln L)$, $\max C_L(M,N)=\Omega(1)$;
 	\item when $M= O(1)$ and $N = O(\ln L)$, $\max C_L(M,N) =\Omega(1/\ln L)$. 
 \end{enumerate}
\end{theorem}
\begin{IEEEproof}%
Let $r = \frac{M}{N}\log |\mc A|$. Substituting the error bound of $\epsilon_{\ell}$ in \eqref{eq:23} into \eqref{eq:22}, we obtain the end-to-end decoding error bound:
\begin{IEEEeqnarray}{rCl} 
	P(g_L(\cv Y_L) \ne \cv X)  & \le &
	1 - 
	\prod_{\ell=1}^L(1-e^{-N\ee_{\ell}(r)
	}) \\
        & \le & 1 - \left(
  1 - e^{-N\ee^*(r)}
  \right)^L. \label{eq:decoding:error:e2e}
\end{IEEEeqnarray}
Using a similar argument as in the proof of \cite[Theorem V.3]{Niesen2007}, the GBNC achieves rate $r\left(
  1 - e^{-N\ee^*(r)}
  \right)^L - 1/N$.

Next, we discuss the scalability of the rate for different scalings of $M$ and $N$.
1) Suppose $M=\Theta(N)$, i.e., $r_1\leq r \leq r_2$ for some $0<r_1<r_2<C'$, and $N=O(\ln L)$.
In this case, 
\begin{equation}
\max C_L(M,N)\ge 
r_1\left(
  1 - e^{-N\ee^*(r_2)}
  \right)^L - \frac{1}{N}.\label{Eq:lower:bound:GBNC}
\end{equation}
Since $N=O(\ln L)$, it holds that $\left(
  1 - e^{-N\ee^*(r_2)}
  \right)^L=\Theta(1)$ and $1/N=o(1)$.
  Consequently, the lower bound in \eqref{Eq:lower:bound:GBNC} is $\Theta(1)$.

  2) Suppose $M=O(1)$, i.e., $ r\leq r_3$ for some $0<r_3<C$ and $N=O(\ln L)$.
Then it holds that $\left(1 - e^{-N\ee^*(r)}\right)^L \geq \left(1 - e^{-N\ee^*(r_3)}\right)^L$.
Similarly, the rate of GBNC is lower bounded by
\begin{equation}
\frac{1}{N}\left( 
M\log|\mc A|\left(
  1 - e^{-N\ee^*(r_3)}
  \right)^L-1
\right).
\label{Eq:lb:case:2}
\end{equation}
Since $N=O(\ln L)$ and $M=O(1)$, one can properly choose these parameters to ensure $\left(
  1 - e^{-N\ee^*(r_3)}
  \right)^L>\frac{1}{M\log|\mc A|}$.
  Consequently, the lower bound in \eqref{Eq:lb:case:2} is $\Theta(1/N)=\Theta(1/\ln L)$.
\end{IEEEproof}

\begin{figure}[t]
     \centering
\includegraphics[width=0.45\textwidth]{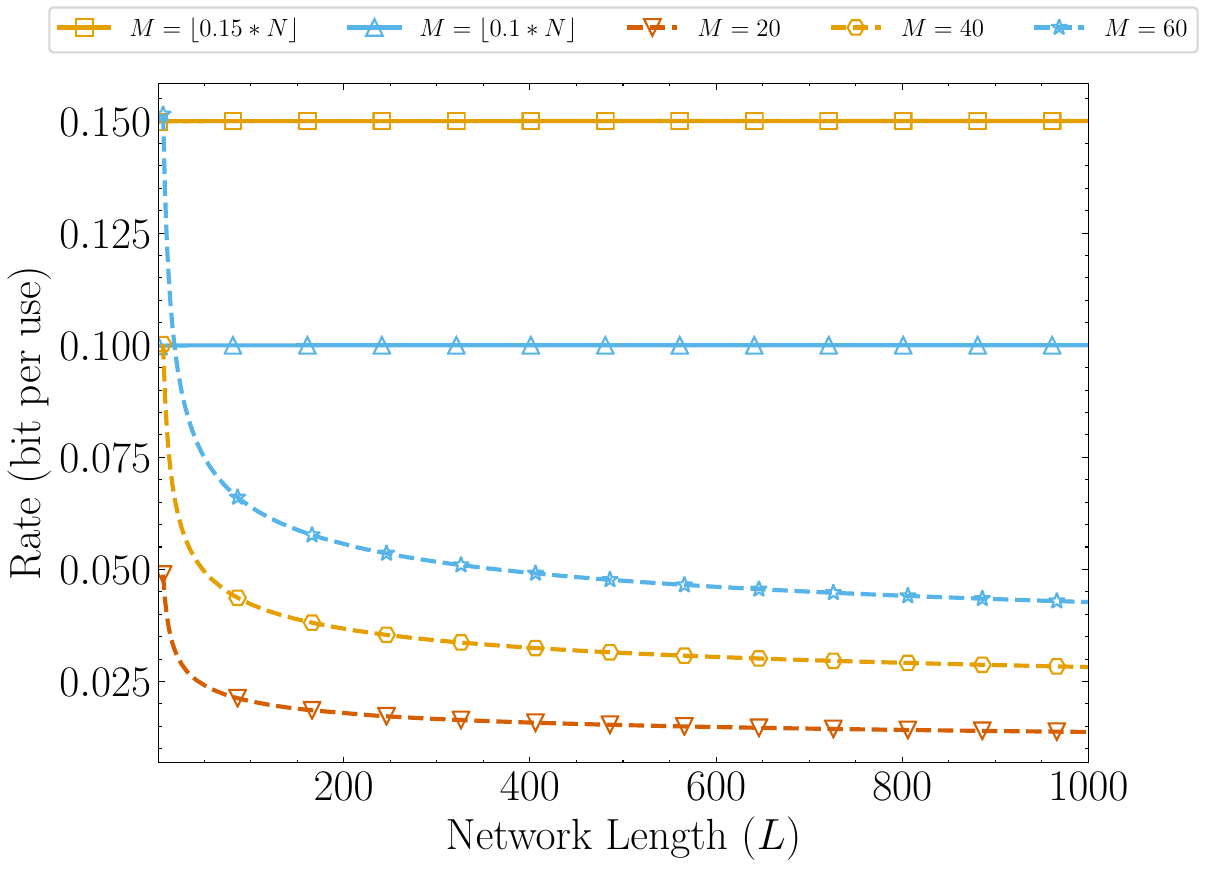}
\caption{
Numerical illustrations of the achievable rates of GBNC based on decode-and-forward recoding~(see \eqref{Eq:lower:bound:CL}) using BSC with crossover probability $\epsilon=0.2, |\mc A|=2$.
The hop length $L$ ranges from $1$ to $1000$ and inner blocklength $N=\lfloor 10^4\log(L+1)\rfloor$.
The solid lines correspond to achievable rates when 
$M=\lfloor c_1N\rfloor$, and the dashed lines correspond to rates when $M=c_1'$.
Here constants $c_1\in\{0.1,0.15\}, c_1'\in\{20,40,60\}$.}
        \label{fig:numerical:BSC:Ga}
\end{figure}

We provide an example showcasing the achievable rates of GBNC based on decode-and-forward in Fig.~\ref{fig:numerical:BSC:Ga}: We use a line network formed by the BSC with crossover probability $\epsilon=0.2$ and vary the number of hops $L$ from $1$ to $1000$, and we use GBNC with $|\mc A|=2$. 
The solid lines correspond to the case with $M=\Theta(N), N=O(\ln L)$, where we observe that the achievable rate remains to be a constant for increasing $L$.
The dash lines correspond to the case with $M=O(1), N=O(\ln L)$, where the achievable rate decays slowly when $L$ increases.

As a summary, decode-and-forward recoding can achieve the same order of rate scalability as the upper bound in Theorem~\ref{thm:scalability} for case 2) and 3), where the buffer size requirement is $O(N)= O(\ln L)$. 
The above approach, however, cannot be used to show the scalability with $M=O(1)$ and $N=O(1)$, since the lower bound in Theorem~\ref{Theorem:lower:bound:decoding:forward} is negative when $N=O(1)$ and $L$ is large.
This case will be discussed in Sec.\ref{sec:reduction} using another approach.

\subsection{Repetition Recoding}
\label{sec:rep}

In this subsection, we show that it is possible to achieve $\Omega(1/\ln L)$ using $M=O(1)$ and $N=O(\ln L)$, while the buffer size requirement is $O(\ln\ln L)$.
Specifically, we discuss the repetition recoding scheme, which is a special decode-and-forward recoding scheme.
In the following, we introduce this recoding scheme by specifying $f_\ell, g_\ell, \ell=1,\ldots,L$ defined in \S\ref{sec:relatedscheme}.

We first discuss the case $M=1$. For any $\ell$, let $\Qin^\ell$ be the {maximal} subset of $\Qin$ such that for any $x\neq x'\in \Qin^\ell$, $Q_{\ell}(\cdot|x) \neq Q_{\ell}(\cdot|x')$. 
For $\ell=1,\ldots,L$, assume $|\Qin^\ell| \geq |\mathcal{A}| \geq 2$, and let $u_\ell$ be a one-to-one mapping from $\mc A$ to $\Qin^\ell$.
For a generic batch $x\in\mathcal{A}$ with $M=1$, node $\ell-1$ transmits $u_\ell(x)$ for $N$ times, i.e.,
\begin{equation}
  \label{eq:f1}
  f_\ell(x) = (u_\ell(x),\ldots,u_\ell(x)).
\end{equation}
Suppose $\cv Y_{\ell} = \cv y_{\ell}$, i.e., node $\ell$ receives $\cv y_{\ell}$ for the transmission $f_\ell(x)$. The decoding function $g_{\ell}$ is defined based on the maximum likelihood~(ML) criterion:
\begin{equation}\label{eq:g1}
g_{\ell}(\cv y_{\ell}) = \argmax_{x\in \mc A}~\prod_{i=1}^{N}Q_{\ell}(\cv y_{\ell}[i]\mid u_{\ell}(x)),
\end{equation}
where a tie is broken arbitrarily. 
Let
\begin{equation}
\begin{aligned}
  \mathcal{L}_{\ell}(x; \cv y_{\ell}) &= 
\sum_{i=1}^{N}\ln Q_{\ell}(\cv y_{\ell}[i]\mid u_{\ell}(x)) \\
&= \sum_{y\in\Qout}~ \mathcal{N}(y|{\cv y_{\ell}})\ln Q_{\ell}(y\mid u_{\ell}(x)),
\end{aligned}
\end{equation}

\noindent where $\mathcal{N}(y|{\cv y_{\ell}})$ denote the number of times that $y$ appears in $\cv y_{\ell}$. 
Then the ML decoding problem can be equivalently written as $g_{\ell}(\cv y_{\ell}) = \argmax_{x\in \mc A}~\mathcal{L}_{\ell}(x; \cv y_{\ell})$.

To perform the ML decoding, node $\ell$ needs to count the frequencies of symbols $y$ for any $y\in\Qout$ among $N$ received symbols. 
As a result, a buffer of size $O(|\Qout|\log N)=O(\ln N)$ at each intermediate is required. Additionally, the computation cost of the repetition recoding is $O(N)$ per batch.
The following lemma bounds the maximum decoding error probability $\epsilon_{\ell}$ of $(f_{\ell},g_{\ell})$ for $Q_{\ell}$.%

\begin{lemma}\label{proposition:decoding:error:M1}
For any $\ell=1,\ldots,L$, under the condition $|\Qin^\ell| \geq |\mathcal{A}|\geq 2$, using the repetition encoding $f_\ell$ and the ML decoding $g_{\ell}$ in \eqref{eq:f1} and \eqref{eq:g1}, respectively, the maximum decoding error probability $\epsilon_{\ell}$ for $Q_{\ell}$ satisfies $\epsilon_{\ell}\le \big(|\mathcal{A}|-1\big)\exp\left(
-NE_{\ell}
\right),$
where $E_{\ell}>0$ is a constant depends only on the channel $Q_{\ell}$.
\end{lemma}

  Consider a sequence of DMCs $Q_{\ell},\ell=1,2,\ldots$ with %
$C(Q_{\ell})>0,\ell\geq 1$. Let
\begin{equation}
  E^* = \inf\{E_{\ell}, \ell\ge1\}.
\end{equation}
We choose the alphabet $\mc A$ such that $|\mc A|\in[2, S^*]$, where
  \begin{equation}
  S^* = \inf\{|\Qin^{\ell}|:~\ell\ge1\}.
\end{equation}
Note that when $C(Q_{\ell})>0$, $|\Qin^{\ell}| \geq 2$. Hence $S^* \geq 2$.
Considering the repetition coding,
\begin{equation*}
  P(g_L(\cv Y_L) = \cv X)  \geq \left(1 - e^{-NE^*}\right)^L.
\end{equation*}
Applying an argument in \cite[Theorem V.3]{Niesen2007}, we obtain the following theorem.
\begin{theorem}\label{Theorem:lower:bound:M1:repetition}
For the line network of length $L$, %
the GBNC with repetition recoding scheme, batch size $M=1$, inner blocklength $N$, and batch alphabet $\mc A$
achieves rate
\begin{equation}\label{Eq:lb:1N}
\begin{aligned}
C_L(1,N) & \ge 
\frac{1}{N}\bigg\{
   \log |\mc A|  - \mathcal{H}\left(\left(1 - e^{-NE^*}\right)^L\right) \\
  & - \left(1 - \left(1 - e^{-NE^*}\right)^L\right)\log (|\mc A|-1)
\bigg\},
\end{aligned}
\end{equation}
where $\mathcal{H}(\cdot)$ denotes the binary entropy function.
When  $N=O(\ln L)$, $\max C_L(1,N)=\Omega(1/\ln L)$.
\end{theorem}

\begin{figure}[t]
     \centering
     \subfigure[the optimal value of $N$, denoted by $N_L^*$, that maximizes the lower bound of $C_L(1,N)$ in \eqref{Eq:lb:1N}]{
    \includegraphics[width=0.4\textwidth]{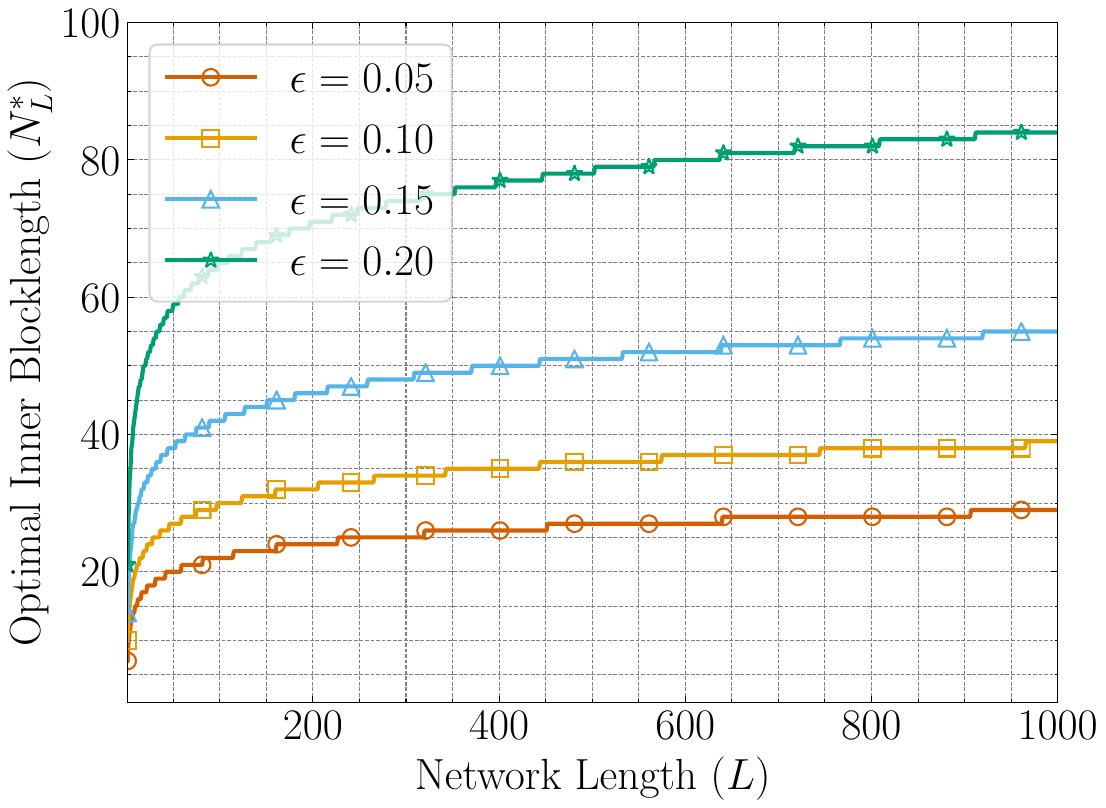}}
      \subfigure[the achievable rates with optimized inner blocklength $N$ and fixed batch size $M=1$]{
    \includegraphics[width=0.4\textwidth]{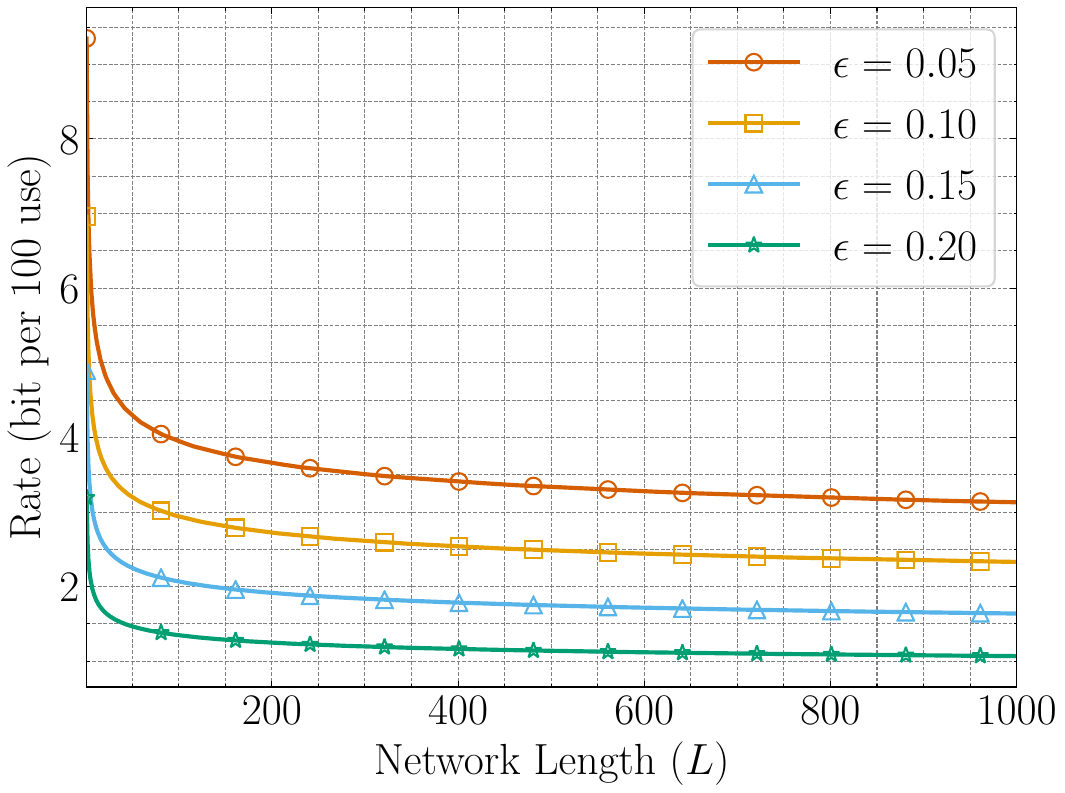}} 
\caption{
Numerical illustrations of the achievable rates of GBNC based on repetition recoding~(see \eqref{Eq:lb:1N}) using BSC with crossover probability $\epsilon=0.05, 0.1, 0.15, 0.2, |\mc A|=2$.
}
        \label{fig:numerical:BSC}
\end{figure}

We plot the rate of repetition recoding using BSC with crossover error probability $\epsilon\in\{0.05, 0.1, 0.15, 0.2\}$ and $|\mc A|=2$ with respect to the hop length $L$ in Fig.~\ref{fig:numerical:BSC}.
In Fig.\ref{fig:numerical:BSC}(a), for each hop length $L$, we plot the optimal value of $N$ maximizing the lower bound~\eqref{Eq:lb:1N}, which is denoted as $N_L^*$.
This illustration highlights the observed trend of $N_L^*$ increasing roughly in the order of $\ln L$.
In Fig.~\ref{fig:numerical:BSC}(b), we plot the lower bound \eqref{Eq:lb:1N} for each hop length $L$, showcasing an approximate decrease rate in the order of $1/\ln L$.

The repetition coding scheme discussed previously has a limitation that $|\mc A| \leq |\Qin|$. We can extend the scheme by multiple uses of $Q_{\ell}$. 
For an integer $m$, let $\Qin^{m, \ell}$ be the maximum subset of $\Qin^{m}$ such that for any $\cv x \neq \cv x' \in \Qin^{m, \ell}$, $Q_{\ell}^{\otimes m}(\cdot|\cv x) \neq Q_{\ell}^{\otimes m}(\cdot|\cv x')$. 
Define
\begin{equation}
S^{m,*} = \inf\{|\Qin^{m, \ell}|:~\ell\ge1\}.
\end{equation}
Fix $m$, $M$ and a finite alphabet $\mc A$ such that $|\mc A|^M\in[2,S^{m,*}]$.
Conseuqently, we can view the line network of channels $Q_1,\ldots, Q_L$ as one of $Q_{1}^{\otimes m}, \ldots, Q_L^{\otimes m}$.
For the latter, we can apply the repetition recoding with batch size $1$, inner blocklength $\tilde{N}$ and the batch alphabet $\mc A^{M}$, which for the original line network of $Q_1,\ldots,Q_L$ is a GBNC with batch size $M$, inner blocklength $m\tilde{N}$ and the batch alphabet $\mc A$. 
Based on Theorem~\ref{Theorem:lower:bound:M1:repetition},
such a coding scheme achieves rate
\begin{equation}
\begin{aligned}
  \frac{1}{\tilde N}\bigg\{ &
  \log |\mc A|^M  - \mathcal{H}\left(
\left(1 - e^{-\tilde NE^{m,*}}\right)^L\right) \\
  & - \left(
1 - \left(1 - e^{-\tilde NE^{m,*}}\right)^L\right)\log (|\mc A|^M-1) 
\bigg\}.
\end{aligned}
\end{equation}

While the repetition code may appear straightforward, it serves as an illustrative example of how to reduce the buffer size at the intermediate node.
Using convolutional codes with Viterbi decoding, due to their analogous encoding and decoding nature, can achieve the same order of the buffer size.
However, the corresponding achievable rate is challenging to analyze.

\subsection{Channel Reduction}
\label{sec:reduction}

When all the links in the line network are identical DMCs, it has been shown in~\cite{Niesen2007} that an exponentially decreasing rate can be achieved using $N=O(1)$, which corresponds to the first case in Theorem~\ref{thm:scalability}. Here we discuss how to generalize this scalability result to line networks where the DMCs $Q_\ell$ are not necessarily identical.
Our approach is to perform recoding so that the line network is reduced to one with identical channels.

We introduce the reduction of an $m\times n$ stochastic matrix $Q$ with $C(Q)>0$.
Let $r=\rank(Q)$. Note that $C(Q)>0$ if and only if $r\geq 2$.
Let $s$ be an integer such that $2\leq s \leq r$. 
We would like to reduce $Q$ by multiplying an $s\times m$ matrix $R$ and an $n\times s$ matrix $S$ before and after $Q$, respectively, so that $RQS$ becomes an $s\times s$ matrix  $U_s(\varrho) $ with $(U_s(\varrho))_{i,j}=\varrho$ if $i=j$ and otherwise $(U_s(\varrho))_{i,j}=\frac{1-\varrho}{s-1}$,
where $\varrho$ is a parameter in the range $(1/s, 1]$.
When $1/s<\varrho\leq 1$, 
among all the $s\times s$ stochastic matrices with trace $s\varrho$, $U_s(\varrho)$ is the one that has the least mutual information for the uniform input distribution~(ref. \cite[Theorem V.3]{Niesen2007}).
The reduction described above, if exists, is called \emph{uniform reduction}. 

We give an example of uniform reduction with $s=2$. Choose $R$ so that $RQ$ is an $s$-row matrix formed by $s$ linearly independent rows of $Q$. Let $a_{ij}$ be the $(i,j)$ entry of $RQ$, where $i=1,2$ and $1\leq j \leq n$.
Define an $n\times 2$ stochastic matrix $W=(w_{ij})$ as
\begin{equation}\label{eq:w}
   w_{i1} =
  \begin{cases}
    \frac{a_{1i}}{a_{1i}+a_{2i}} & \text{ if } a_{1i}+a_{2i}>0, \\
    1 & \text{ otherwise},
  \end{cases}
\end{equation}
and $w_{i2}=1-w_{i1}$, where $1\le i\le n$.
With the above $R$ and $W$, we see that $  RQW  = U_2(\varrho)$,
where $\varrho = \sum_{k:a_{1k}+a_{2k}>0} \frac{a_{1k}^2}{a_{1k}+a_{2k}}$. The following lemma states a range of $\varrho$ such that the reduction to $U_2(\varrho)$ is feasible.

\begin{lemma}\label{lm:r2}
  For a stochastic matrix $Q$ such that $C(Q)>\epsilon$ for some $\epsilon>0$, 
  there exists a constant $B>1/2$ depending only on $\epsilon$ such that $Q$ has a uniform reduction to $U_2(\varrho)$ for all $1/2 < \varrho \leq B$.   
\end{lemma}

Fix any $\epsilon>0$.
Consider the line network formed by $Q_1,\ldots,Q_L$, where $C(Q_{\ell})>\epsilon$ and hence $\rank(Q_{\ell})\geq 2$. We discuss a GBNC with $|\mc A| = 2$ and $M=N=1$. By Lemma~\ref{lm:r2}, there exists $\varrho>1/2$ such that for any $\ell$, there exists stochastic matrices $R_{\ell}$ and $S_{\ell}$ such that $R_\ell Q_\ell S_\ell = U_2(\varrho)$.
Define the recoding at the source node as $R_{1}$, and for $\ell=1,\ldots,L-1$, define the recoding at node $\ell$ as $S_{\ell}R_{\ell+1}$. At the destination node, process all the received batches by $R_L$. The overall operation of a batch from the source node to the destination node is $W_L'\triangleq (U_2(\varrho))^{L}$. Applying the argument in \cite[Theorem~III.5]{Niesen2007}, we get
\begin{IEEEeqnarray}{rCl}
\log
\Big(
\frac{1}{2\varrho - 1}\Big)
&\le&
\liminf\limits_{L\to\infty}-\frac{1}{L}\log C(W_L')\\
&\le& 
\limsup\limits_{L\to\infty}-\frac{1}{L}\log C(W_L') \\
&\le & 
2\log
\Big(\frac{1}{2\varrho - 1}\Big), 
\end{IEEEeqnarray}
where $\frac{1}{2\varrho - 1}$ is the second largest eigenvalue of $U_2(\varrho)$.
Therefore, a channel code for the transition matrix $W_L$ as the outer code can achieve the rate $\Omega(e^{-cL})$ as $L\to\infty$, where the constant $c$ is between $\log\left(
\frac{1}{2\varrho - 1} \right)$ and $2\log\left(
\frac{1}{2\varrho - 1} \right)$. The above discussion is summarized as the following theorem:

\begin{theorem}\label{thm:xd2}
  Consider a sequence of DMCs $Q_{\ell},\ell=1,2,\ldots$ with $\inf\{C(Q_{\ell}),\ell\geq 1\}>0$. 
  For the line network of length $L$, where the $\ell$th link is $Q_{\ell}$, the GBNC with $M=O(1)$ and $N=O(1)$ achieves rate
$
C_L(M,N)\ge c'\cdot e^{-cL},
$
where $c$ is a constant between $\log\left(
\frac{1}{2\varrho - 1} \right)$ and $2\log\left(
\frac{1}{2\varrho - 1} \right)$, and $c'>0$ is a constant.
\end{theorem}

The technique used in the proof of Theorem~\ref{thm:xd2} can be generalized for $M,N\geq 1$.
We first show that for an $m\times n$ stochastic matrix $Q$ with $\rank(Q)\geq 2$, for any $2\leq s\leq r$, the uniform reduction to $U_s(\varrho)$ exists if $\varrho$ is sufficiently close to $1/s$.
For an integer $2 \leq s\leq r$, let 
\begin{equation}
  \kappa_s(Q) = \max_{\substack{s\times m \text{ stochastic matrix }  R\\ n\times s\text{ stochastic matrix }W}} \min\mathrm{inv}(RQW)
\end{equation}
where $\min\mathrm{inv}(RQW)$ is the minimum value of $(RQW)^{-1}$ when $RQW$ is invertible and is $\infty$ otherwise.
We give an example of $R$ and $W$ such that $RQW$ is invertible. Choose $R$ so that $RQ$ is an $s$-row matrix formed by $s$ linearly independent rows of $Q$. Let $a_{ij}$ be the $(i,j)$ entry of $RQ$, where $1\leq i \leq s$ and $1\leq j \leq n$.
To simplify the discussion, we assume all the columns of $RQ$ are non-zero.
Define $W = D (RQ)^{\top},$
where $D$ is an $n\times n$ diagonal matrix with the $(i,i)$ entry $1/\sum_{j'}a_{j'i}$. With the above $R$ and $W$, we see that $RQW$ is positive definite and hence invertible.
Let $  \rho_s(Q) = \frac{\min\{\kappa_s(Q),0\} -1}{s\min\{\kappa_s(Q),0\} -1}.$
We see that $\rho_s(Q)>1/s$.  
The following lemma states a range of $\varrho$ such that the reduction to $U_s(\varrho)$ is feasible. 

\begin{lemma}\label{lemma:red1}
  Consider an $m\times n$ stochastic matrix $Q$ with rank $r\geq 2$.
  For any $2\leq s \leq r$ and $1/s < \varrho \leq \rho_s(Q)$, there exist an $s\times m$ stochastic matrix $R$ and an $n\times s$ stochastic matrix $S$ such that $RQS = U_s(\varrho)$.  
\end{lemma}

\begin{remark}
  Lemma~\ref{lm:r2} is stronger than Lemma~\ref{lemma:red1} for the case $s=2$ as the former gives a uniform bound on $B$ that does not depend on $Q$ as long as $C(Q)> \epsilon$. 
\end{remark}

Consider a line network formed by $Q_1,\ldots,Q_L$, where $C(Q_{\ell})>0$ and hence $\rank(Q_{\ell})\geq 2$. Let $r = \min_{\ell=1}^L \rank(Q_{\ell})$. Assuming $r\geq |\mc A|$, we first discuss a recoding scheme with $M=N=1$. Let $\varrho = \min_{\ell=1}^L \rho_r(Q_{\ell})$. By Lemma~\ref{lemma:red1}, there exists stochastic matrices $R_{\ell}$ and $S_{\ell}$ such that $R_\ell Q_\ell S_\ell = U_r(\varrho)$. The following argument is similar as that of the proof of Theorem~\ref{thm:xd2}. 
Now we consider recoding with $M,N = O(1)$.
Fix $M,N = O(1)$ and a finite alphabet $\mc A$ such that $r^N\ge |\mathcal{A}|^M$. Regarding the line network $\mc L$ as one formed by $Q_1^{\otimes N},\ldots, Q_L^{\otimes N}$, we can apply the above GBNC with 
batch size $1$, 
inner blocklength $1$ and the batch alphabet $\mc A^M$, which for the original line network $\mc L$ of $Q_1,\ldots,Q_L$ is a GBNC with batch size $M$, inner blocklength $N$ and {the} batch alphabet $\mc A$.

\section{Line Networks of Packet Erasure Channels}
\label{sec:era}

For line networks of \emph{packet erasure channels}, GBNC is also
called batched network coding (BNC).  In this section, we discuss line
networks with identical {packet erasure channels}, for which, we
demonstrate stronger converse and achievability results than the
general ones.

Fix the alphabet $\mc A$ with $|\mc A|\geq 2$.  Suppose that
the input alphabet $\Qin$  and the output alphabet $\Qout$ are both $\mc A \cup \{e\}$ where $e\notin \mc A$ is called the erasure. For example, we may use a sequence of bits to represent a packet so that $\mc A = \{0,1\}^T$, i.e., each packet is a sequence of $T$ bits. Henceforth, a symbol in $\mc A$ is also called a packet in this section.
A packet erasure channel with erasure probability $\epsilon$ ($0<\epsilon<1$) has 
the transition matrix $Q_{\text{era}}$: for each $x \in \mc A$, $Q_{\text{era}}(y|x)=1-\epsilon$ if $y=x$ and  $Q_{\text{era}}(y|x)=\epsilon$ if $y=e$.
The input $e$ can be used to model the input when the channel is not used for transmission and we define $Q_{\text{era}}(e|e) = 1$. When the input $e$ is not used for encoding information, erasure codes can achieve a rate of $1-\epsilon$ symbols (in $\mc A$)  per use. It is also clear that $C_0(Q_{\text{era}})=0$.

\subsection{Upper Bound}
We obtain a refined upper bound by using a simpler channel function
for packet erasure channels: The relation between the input $X$ and
output $Y$ of a packet erasure channel can be written as a function
\begin{equation}\label{eq:aera}
  Y = \alpha_{\text{era}}(X,Z) =
  \begin{cases}
    X & \text{if } Z \neq e, \\
    e & \text{if } Z = e,
  \end{cases}
\end{equation}
where $Z$ is a discrete random variable independent of $X$ with $P(Z = e) = \epsilon$. In other words, $Z$ indicates whether the channel output is the erasure or not.

For a line network of length $L$ with a GBNC of inner blocklength $N$, the bottleneck status can be defined as
\begin{equation}
  \label{eq:9}
  E_0 = \{\lor_{\ell=1}^L(\cv Z_\ell[i] = e, i = 1,\ldots,N)\},
\end{equation}
where $\cv Z_{\ell}[i]$ is the channel variable of the $i$th use of $Q_{\ell}$.
With this bottleneck status, $p_1=P(\overline{E_0})=(1-\epsilon^N)^L$.
Following a similar procedure as in the proof of Theorem~\ref{thm:ca}, we have
\begin{equation}
C_L(M,N)\le \frac{(1-\epsilon^N)^L}{N} \min\{ M \log|\mc A|, N\log |\Qout| \},
\label{Eq:upper:line:net:packet}
\end{equation}
which is a tighter upper bound than \eqref{eq:1ca}.

\subsection{Achievability by Random Linear Recoding}

We now introduce a class of inner codes with batch size $M= O(1)$, which provides the achievability counterpart for the cases 1) and 2) in Theorem~\ref{thm:ca}. Let $\ff_q$ be the finite field of $q$ symbols, and let $T>0$ be an integer. Suppose $\mc A = \ff_q^T$, i.e., each packet is a sequence of $T$ symbols from the finite field $\ff_q$.
The outer code generates batches that consist of $M$ packets in $\mc A$, and can be represented as a $T\times M$ matrix over $\ff_q$.
In each packet generated by the outer code, the first $M$ symbols in $\ff_q$ are called the \emph{coefficient vector}. 
A batch $\cv X$ has the first $M$ rows, called the \emph{coefficient matrix}, forming the identity matrix. In the following discussion, we treat the erasure $e$ as the all-zero vector $\cv 0$ in $\ff_q^T$, which is not used as a packet in the batches. 
In other words, when a packet is erased, an intermediate node assumes $\cv 0$ is received. 

The inner code is formed by \emph{random linear recoding}, which have been studied in random linear network coding (RLNC).
A random linear combination of vectors in $\mc A$ has the linear combination coefficients chosen uniformly at random from $\ff_q$. The inner code includes the following operations: 
\begin{itemize}
\item The source node generates $N$ packets for a batch using random linear combinations of the $M$ packets of the batch generated by the outer code.
\item Each intermediate node generates $N$ packets for a batch using random linear combinations of all packets of the received packets of the batch.
\end{itemize}
Note that for each batch, only the packets with linearly independent coefficient vectors are needed for random linear recoding. Therefore, the buffer size used to store batch content is $O(MT\log q)$ bits.
Also, the computational cost of the above recoding scheme for each intermediate node is $O(N^2T\log q)$ per batch.

At each node, the rank of the coefficient matrix of a batch (i.e., the first $M$ rows of the matrix formed by the generated/received packets of the batch) is also called the rank of the batch. At each node, the ranks of all the batches follow an identical and independent distribution.
Denote by $\pi_\ell$ the rank distribution of a batch at node ${\ell}$. 
As all the batches at the source node have rank $M$, we know that $\pi_0=(0,0,\ldots,0,1)$.
Moreover, the rank distributions $\pi_0,\pi_1,\ldots, \pi_L$ form a Markov chain so that for $\ell=1,\ldots, L$, it holds that
\begin{equation}\label{eq:tran:rk}
  \pi_{\ell} = \pi_{\ell-1} \cv P 
\end{equation}
where $\cv P$ is the transition matrix characterized in \cite[Lemma 4.2]{yang17monograph}. %

\begin{figure}[t]
     \centering
\subfigure[plot of $\text{BNC}_L(M,N)$ and $\text{PEC}^{\text{UB}}_L(M,N)$ when $L$ increases for the case $M=N$ fixed]{
\includegraphics[width=0.4\textwidth]{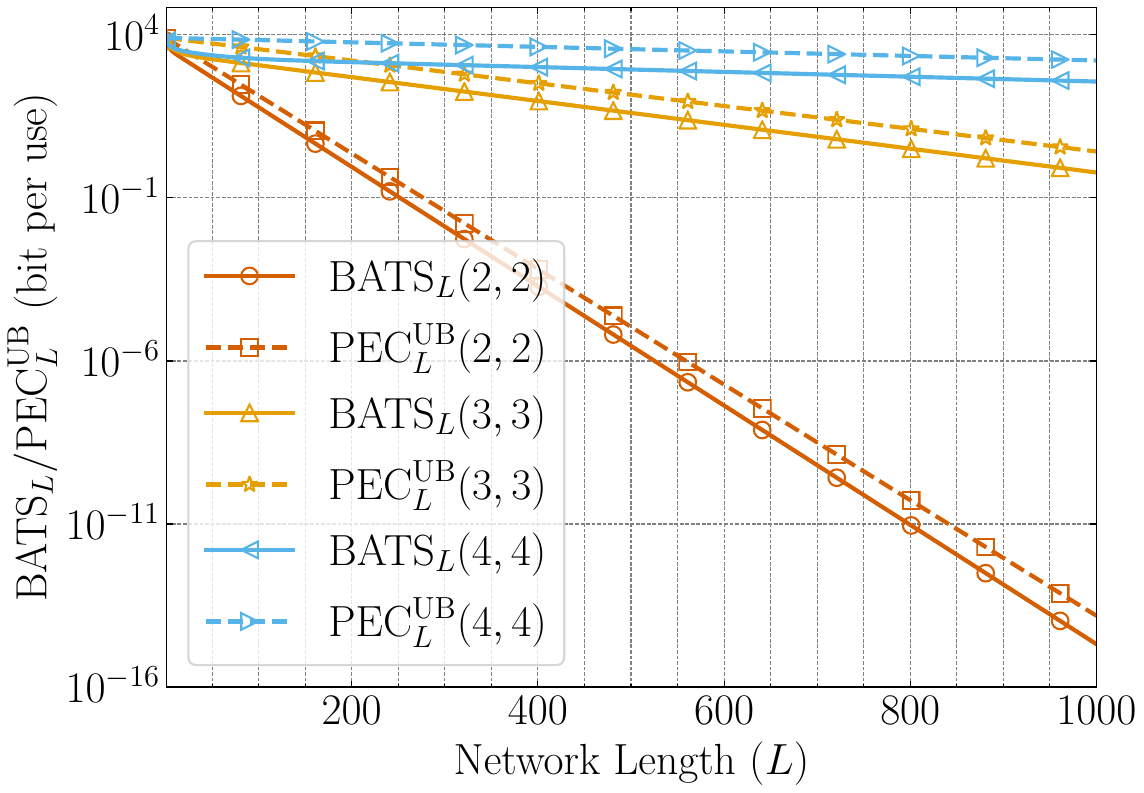}}
    \subfigure[plot of the optimal value of $N$, denoted by $N_L^*$, that maximizes $\text{BNC}_L(M,N)$ for a fixed value of $M$]{
\includegraphics[width=0.4\textwidth]{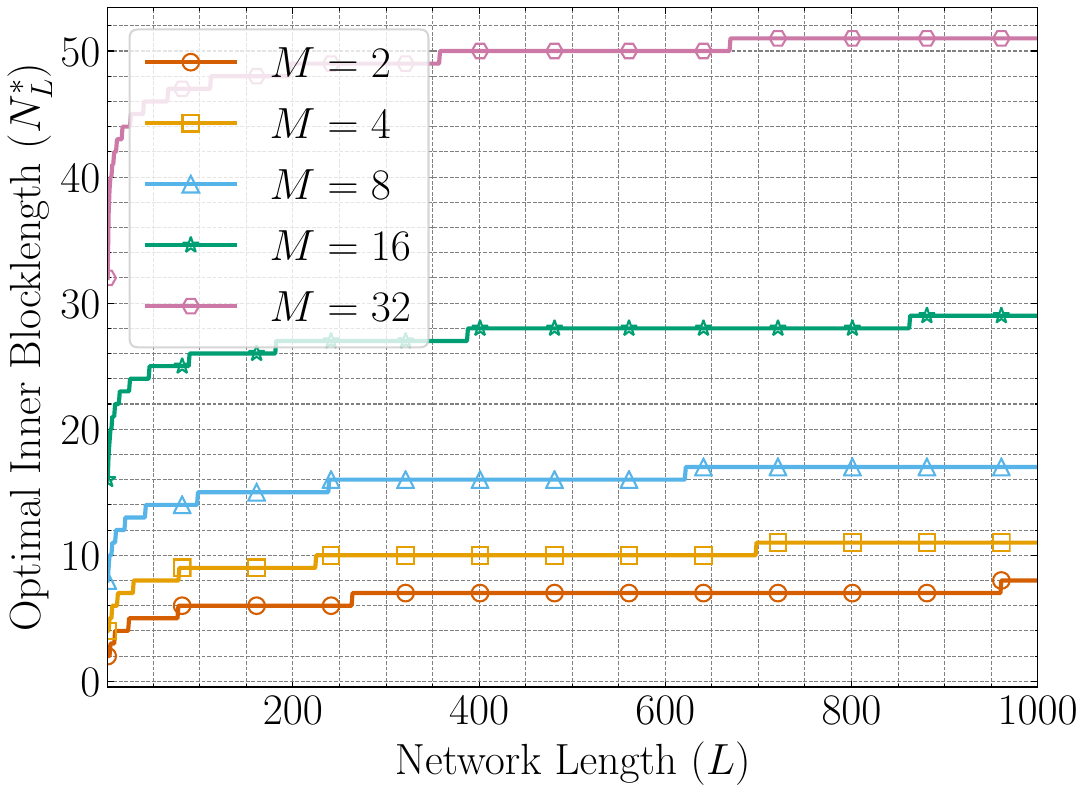}}
       \subfigure[plot of $\text{BNC}_L(M,N)$ and $\text{PEC}^{\text{UB}}_L(M,N)$ when $L$ increases for the case that $M$ is fixed and $N=N_L^*$]{
\includegraphics[width=0.44\textwidth]{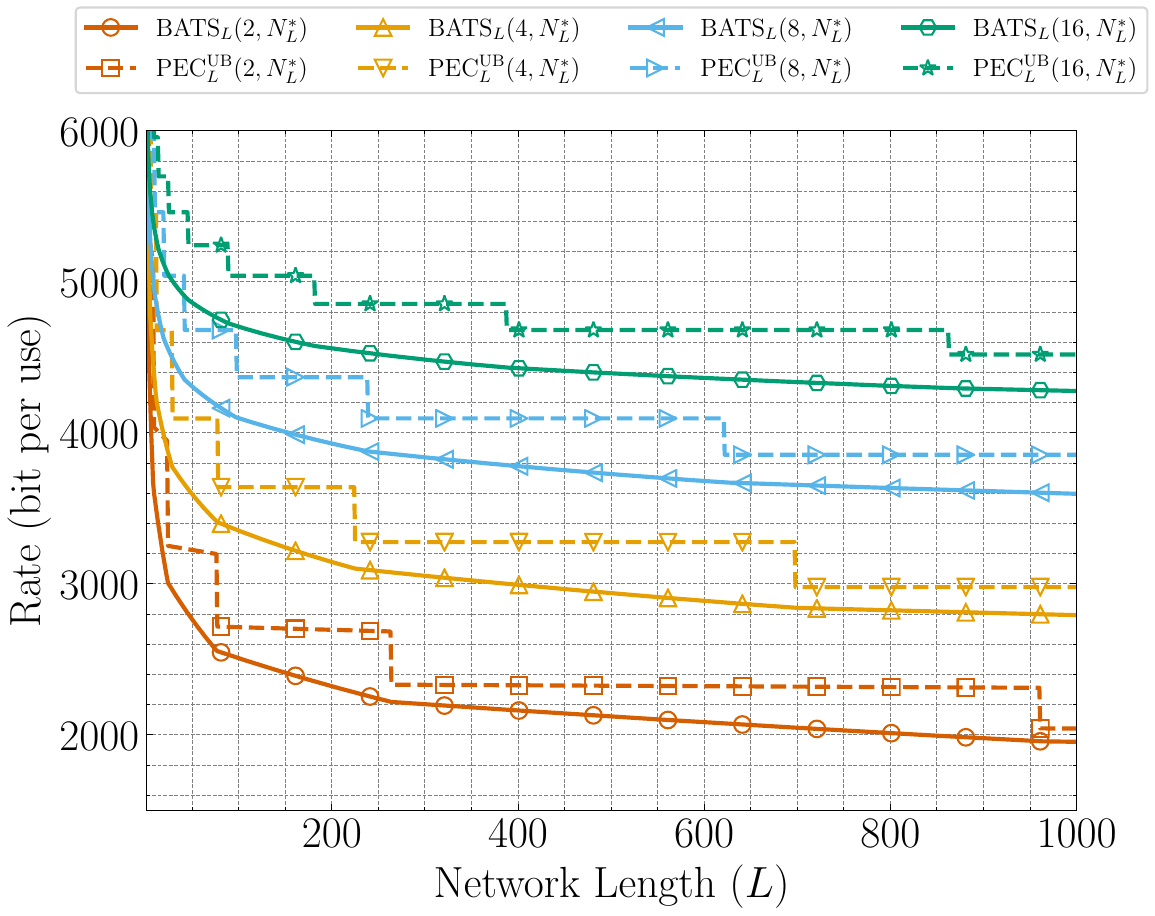}}
\caption{
{Numerical illustrations of the upper bound and achievable rates of BNC.
   }
        }
        \label{fig:numerical}
\end{figure}

The maximum achievable rate of this class of BNC is $(1-\frac{M}{T})\frac{\E[\pi_L]}{N}$ packets (in $\mc A$) per use, and can be achieved by BATS codes~\cite{yang10bf,yang17monograph}, where the factor $1-\frac{M}{T}$ comes from the overhead of $M$ symbols in a packet used to transmit the coefficient vector. Denote 
\begin{equation}
  \label{eq:batsrate}
  \text{BNC}_L(M,N) = \left(1-\frac{M}{T}\right)\frac{\E[\pi_L]}{N}\log|\mc A|.
\end{equation}
In Fig.~\ref{fig:numerical}, 
we compare numerically the upper bound and the achievable rates of BNC by evaluating \eqref{Eq:upper:line:net:packet} and \eqref{eq:batsrate}, respectively.
{Throughout the experiment, we specify parameters $\epsilon=0.2$, $q=256$ and $T=1024$ following the same setup as in \cite[Fig.~10]{yang14bats}, which are decided based on the following considerations: Firstly, in many practical wireless communication systems, a packet loss rate of around 10 to 20 percent is commonly observed. 
Secondly, a finite field of size $256$ is frequently utilized in real-world implementations. 
Lastly, a packet of $1024$ bytes is a typical choice in internet-based communication scenarios. 
}
Note that each packet has $8192$ bits and the min-cut is $6553.6$ bits per use.

First, we consider fixed $M=N=2,3,4$, and plot the calculation for $L$ up to $1000$ in Fig.~\ref{fig:numerical}(a). We see from the figure that for a fixed $N$, the achievable rates of BNC and the upper bound in \eqref{Eq:upper:line:net:packet} share the same exponential decreasing trend.

Second, we consider fixed $M=2,4,8,16,32$. For each value of $M$, we find the optimal value of $N$, denoted by $N_L^*$, that maximizes $\text{BNC}_L(M,N)$. We see from Fig.~\ref{fig:numerical}(b) that $N_L^*$ demonstrates a low increasing rate with $L$. We further illustrate $\text{BNC}_L(M,N^*_L)$ and $\text{PEC}^{\text{UB}}_L(M,N^*_L)$ for each value of $M$ in Fig.~\ref{fig:numerical}(c).

The following theorem justifies the scalability of $\text{BNC}_L(M,N)$ when $L$ is large, where the $M=1$ case was proved in \cite{yang17monograph}.

\begin{theorem}\label{thm:bats}
  Consider a line network of $L$ packet erasure channels with erasure probability $\epsilon$. For GBNCs of fixed batch size $M<T$ and inner blocklength $N$ using random linear recoding,
  \begin{equation}
    \text{BNC}_L(M,N) = \Theta\left(\frac{\left(1-(\epsilon + (1-\epsilon)/q)^N\right)^L}{N}\right).
  \end{equation}
\end{theorem}
When $q$ is relatively large, $\text{BNC}_L(M,N)$ has nearly the same scalability as $\text{PEC}^{\text{UB}}_L(M,N)$, as illustrated by Fig.~\ref{fig:numerical}(c).
Consider two cases of $N$ for the scalability of $\text{BNC}_L(M,N)$:
When $N$ is a fixed number, $\text{BNC}_L(M,N)$ decreases exponentially with $L$. 
When $M$ is a fixed number and $N$ is unconstrained, based on the optimization theory (see, e.g., \cite[Lemma~1]{yang2019capacity}) we know that $\max_N\text{BNC}_L(M,N) = \Theta(1/\ln L),$ and the maximum is achieved by $N= \Theta(\ln L)$. %

\section{Line Networks with Channels of Positive Zero-Error Capacity}
\label{sec:ex}

Last, we discuss how to extend our study so far to line networks of channels that have positive capacity but may also have positive zero-error capacity.
Denote by $\mc L$ a line network of length $L$ formed by channels $Q_1,\ldots,Q_L$, where it is not necessary that $C_0(Q_{\ell})=0$.
For a GBNC on $\mc L$,
the end-to-end transition matrix of a batch is denoted by $W_{\mc L}$. 
Denote the maximum achievable rate of all recoding schemes with batch size $ M $ and inner blocklength $ N $ for $\mc L$ as $ C_{\mc L}(M,N) $. 
Let $L_0$ be the number of channels in $\mc L$ with $ 0 $ zero-error capacity, i.e.,
$L_0=|\{1\leq\ell\leq L: C_0(Q_{\ell})=0\}|.$
In the following{,} we argue that $C_{\mc L}(M,N)$ scales like a line network of length $L_0$ formed by channels with $0$ zero-error capacity.

Let $ \{l_1,\ldots,l_{L_0}\}=\{1\leq\ell\leq L: C_0(Q_{\ell})=0 \} $ where $l_1<l_2<\cdots<l_{L_0}$. Denote by $\mc L'$ the line network formed by the concatenation of $Q_{l_1},\ldots,Q_{L_0}$. %
For any given GBNC on $\mc L$, we can find proper recoding operations for the GBNC on $\mc L'$ so that 
$  W_{\mc L'}=W_{\mc L} $, and hence $ C_{\mc L}(M,N) \leq C_{\mc L'}(M,N)$.
For network $\mc L'$, \S\ref{sec:conv} provides the upper bounds on the achievable rates as functions of length $L_0$ under certain coding parameter sets, which are also upper bounds for network $\mc L$.

We derive a lower bound of achievable rates of $ \mc L$  using the uniform reduction approach introduced in \S\ref{sec:reduction}.
Suppose $C = \inf\{C(Q_{l_i}):~i\geq1\}>0$. 
By Lemma~\ref{lm:r2}, there exists a constant $B\in(1/2,1)$ depending only on $C$ such that there exist stochastic matrices $R_{l_i}$ and $S_{l_i}$ with $R_{l_i}Q_{l_i}S_{l_i} = U_2(B)$ for all $i$. 
For $Q_{\ell}$ with $C_0(Q_{\ell}) >0$, we can find $R_{\ell}$ and $S_{\ell}$ so that $R_{\ell}Q_{\ell}S_{\ell}$ equals the identity matrix $I_2$.
The existence of $R_{\ell}$ and $S_{\ell}$ is guaranteed by the following lemma.%
\begin{lemma}\label{lm:c0}
  For an $m\times n$ stochastic matrix $Q$ with $C_0(Q)>0$, there exists a $2\times m$ stochastic matrix $R$ and a $n\times 2$ stochastic matrix $S$ such that $RQS = I_2$, the $2\times 2$ identity matrix.
\end{lemma}
\begin{IEEEproof}%
For a DMC $Q$, two channel inputs $x_1$ and $x_2$ are said to be \emph{adjacent} if there exists an output $y$ such that $Q(y|x_1)Q(y|x_2)>0$.
Denote by $M_0(Q)$ the largest number of inputs in which adjacent pairs do not exist.
For a DMC with $C_0>0$, we have that $M_0(Q)\geq 2$ and then $C_0(Q)\geq 1$, since otherwise it is easy to verify $ M_0(Q^{\otimes n})\leq 1$ for any $n$ which leads to $C_0(Q) = 0$.

When the channel $Q$ satisfies $C_0(Q)>0$, we have $ M_0(Q)\geq 2 $. Define $R$ as a two-row deterministic stochastic matrix that selects two rows of $Q$ that correspond to two non-adjacent inputs. Denote by $a_{ij}$ the $(i,j)$ entry of $RQ$. We have $a_{1j}a_{2j} = 0$ for all $j=1,\ldots,n$. Let $S$ be defined same as the matrix $W$ in defined in \eqref{eq:w}. 
\end{IEEEproof}

Denote by $\mc L''$ the line network formed by the concatenation of $L_0$ identical channels $U_2(B)$. 
Hence, we obtain $C_{\mc L}(M,N)\geq C_{\mc L''}(M,N)$, where the later can be lower bounded by the techniques in \S\ref{sec:ach}. 
In particular, the error exponent condition in Theorem~\ref{Theorem:lower:bound:M1:repetition} can be verified by checking the proof of Lemma~\ref{proposition:decoding:error:M1} for the special case of BSCs.

\section{Concluding Remarks}

This paper examines the achievable rates of generalized batched network codes (GBNCs) in line networks with general discrete memoryless channels (DMCs). 
The findings suggest that capacity-achieving codes for DMCs may not be the only consideration for the inner code. Simple codes like repetition and convolutional codes can achieve the same rate order while requiring lower buffer sizes. Additionally, reliable hop-by-hop communication is not always optimal when buffer size and latency constraints are present. 

{Feedback is useful in certain communication scenarios. Hop-by-hop feedback does not increase the network capacity (the min-cut). However, exploring its potential benefits is an intriguing area of research in the context of GBNC. Hop-by-hop feedback within batches does not increase the upper bound since it does not increase the capacity of a DMC. However, when hop-by-hop feedback crosses batches, it introduces memory in the batched channel, which may increase the capacity. Additionally, feedback can also simplify coding schemes.}

Future research directions also include investigating better upper bounds and recoding schemes for line networks with special channels like BSCs, generalizing the analysis to channels with infinite alphabets and continuous channels, and exploring whether the upper bound holds for more general codes beyond GBNCs would be valuable.

\newpage
\setcounter{page}{1}
\pagestyle{plain}
\onecolumn
{\centering\LARGE Supplementary Material for ``\it{On Achievable Rates of Line Networks with Generalized Batched
  Network Coding}''\par}
\appendices

\section{Proofs about Converse}
\label{app:proofs:gen}

\begin{IEEEproof}[Proof of Lemma~\ref{lemma:3}]
Write $E_0 = \lor_{\ell=1}^L (E_{0,\ell} \land_{\ell'>\ell}\overline{E_{0,\ell'}})$, where $(E_{0,\ell} \land_{\ell'>\ell}\overline{E_{0,\ell'}})$, $\ell=1,\ldots,L$ are disjoint.
Hence,
\begin{IEEEeqnarray}{rCl}
  P(\cv y_L,\cv x,E_0) & = &  \sum_{\ell=1}^L P(\cv y_L,\cv x, E_{0,\ell},\land_{\ell'>\ell}\overline{E_{0,\ell'}}) 
  \\
  &=&  
 \sum_{\ell} \sum_{ \cv y_{\ell}, \cv u_{\ell}}P(\cv y_L,\cv x,E_{0,\ell}, \land_{\ell'>\ell}\overline{E_{0,\ell'}}, \cv y_{\ell}, \cv u_{\ell})\\
 &= &\sum_{\ell, \cv y_{\ell}}P(\cv y_L, \land_{\ell'>\ell}\overline{E_{0,\ell'}}\mid \cv y_{\ell})
 \sum_{\cv u_{\ell}}P(\cv x,E_{0,\ell}, \cv y_{\ell}, \cv u_{\ell}), \label{eq:l11}
\end{IEEEeqnarray}
where \eqref{eq:l11} from the Markov chain in \eqref{eq:1}.
Further,
\begin{IEEEeqnarray}{rCl}
{\sum_{\cv u_{\ell}}P(\cv x,E_{0,\ell}, \cv y_{\ell}, \cv u_{\ell})} 
& = & \sum_{\cv u_{\ell}}
P(\cv x,\cv u_{\ell})P(E_{0,\ell})P(\cv y_{\ell}\mid \cv u_{\ell}, E_{0,\ell}) \label{eq:l13} \\
& = & \sum_{\cv u_{\ell}}
P(\cv x,\cv u_{\ell})P(E_{0,\ell})P(\cv y_{\ell}\mid E_{0,\ell}) \label{eq:8s9s}
\\
 &=&  p_{\cv X}(\cv x) P(E_{0,\ell})P(\cv y_{\ell}\mid E_{0,\ell}), \label{eq:l12}
\end{IEEEeqnarray}
where \eqref{eq:l13} follows from the channel law, and \eqref{eq:8s9s} follows that $\cv y_{\ell}[i] = y_{\ell}^*$ under the condition $E_{0,\ell}$. 
By \eqref{eq:l11} and \eqref{eq:l12},
\begin{IEEEeqnarray}{rCl}
{P(\cv y_L,\cv x,E_0)}
& = & p_{\cv X}(\cv x)
\sum_{\ell, \cv y_{\ell}}P(\cv y_L, \land_{\ell'>\ell}\overline{E_{0,\ell'}}\mid \cv y_{\ell})P(\cv y_{\ell}, E_{0,\ell}) \\
& = & p_{\cv X}(\cv x)
\sum_{\ell, \cv y_{\ell}}P(\cv y_L, \land_{\ell'>\ell}\overline{E_{0,\ell'}}, \cv y_{\ell}, E_{0,\ell})\\ 
&=& p_{\cv X}(\cv x)P(\cv y_L,E_0),
\end{IEEEeqnarray}
which implies $P(\cv y_L|\cv x, E_0) = P(\cv y_L|E_0)$ and hence $I(p_{\cv X}, W_L^{(0)}) = 0$.
\end{IEEEproof}

\begin{IEEEproof}[Proof of Lemma~\ref{lemma:w1}]
  As $\overline{E_0} = \land_{\ell=1}^L\overline{E_{0,\ell}}$, by \eqref{eq:e0l}, $
P(\overline{E_0}) =  \prod_{\ell=1}^L(
1 - P(E_{0,\ell}) )\leq (1-\varepsilon^{|\Qin|N})^L$. 
We first show that given $\overline{E_0}$, $\cv Z_{1}, \ldots, \cv Z_{L}$ are independent. Write
  \begin{IEEEeqnarray}{rCl}
    {P( \cv z_\ell, \ell = 1,\ldots,L, \overline{E_0})} 
    & = & P(\cv z_\ell, \overline{E_{0,\ell}}, \ell = 1,\ldots,L) \quad  \\
    & = & \prod_{\ell=1}^L P(\cv z_\ell, \overline{E_{0,\ell}}) \\
    &=& \prod_{\ell=1}^L P(\cv z_\ell \mid \overline{E_{0,\ell}}) P(\overline{E_{0,\ell}}). 
  \end{IEEEeqnarray}
  Hence,
  \begin{IEEEeqnarray}{rCl}
    P(\cv z_\ell, \ell = 1,\ldots,L \mid \overline{E_0})
    &=& \prod_{\ell=1}^L P(\cv z_\ell \mid \overline{E_{0,\ell}}) \\
    & = & \prod_{\ell=1}^L P(\cv z_\ell \mid \overline{E_{0}}).
  \end{IEEEeqnarray}
  Under the condition of  $\overline{E_0}$, 
  as $\cv Z_{1}, \ldots, \cv Z_{L}$ are independent,  we have the Markov chain in \eqref{eq:1} holds and hence $I(p_{\cv X}, W_L^{(1)}) \leq I(\cv U_\ell;\cv Y_{\ell} \mid \overline{E_{0,\ell}})$.
\end{IEEEproof}

\begin{IEEEproof}[Proof of Lemma~\ref{lemma:upb}]
  Denote by $\cv y^* = (y^*\cdots y^*)$. We have
  \begin{equation}
    \label{eq:5}
    W(\cv y|\cv x) =
    \begin{cases}
      \frac{Q^{\otimes N}(\cv y^*| \cv x) - p_0}{1-p_0} & \cv y = \cv y^*, \\
      \frac{Q^{\otimes N}(\cv y| \cv x)}{1-p_0} & \text{otherwise}.
    \end{cases}
  \end{equation}
  Let $P(\cv y) = \sum_{\cv x} Q^{\otimes N}(\cv y|\cv x) p(\cv x)$ and $P'(\cv y) = \sum_{\cv x} W(\cv y|\cv x) p(\cv x)$. We have
  \begin{equation}
    \label{eq:6}
    P'(\cv y) =
    \begin{cases}
      \frac{1}{1-p_0} (P(\cv y) - p_0) &  \cv y = \cv y^*, \\
      \frac{1}{1-p_0} P(\cv y) & \text{otherwise}.
    \end{cases}
  \end{equation}
  Substituting \eqref{eq:5} and \eqref{eq:6} into $I(p,W)$, we get
  \begin{IEEEeqnarray}{rCl}
    I(p, W) & = & \sum_{\cv x} p(\cv x) \sum_{\cv y} W(\cv y|\cv x) \log \frac{W(\cv y|\cv x)}{P'(\cv y)} \\
    & = &  \frac{1}{1-p_0} I(p,Q^{\otimes N}) + \frac{1}{1-p_0} U(\cv y^*), \label{eq:up1}
  \end{IEEEeqnarray}
  where
  \begin{IEEEeqnarray}{rCl}
    U(\cv y^*) & = &  \sum_{\cv x} p(\cv x) 
    \bigg((Q^{\otimes N}(\cv y^*|\cv x)-p_0) \log \frac{Q^{\otimes N}(\cv y^*|\cv x)-p_0}{P(\cv y^*)-p_0} \nonumber  \\ & & -Q^{\otimes N}(\cv y^*|\cv x) \log \frac{Q^{\otimes N}(\cv y^*|\cv x)}{P(\cv y^*)} \bigg).
  \end{IEEEeqnarray}
  Using $P(\cv y^*) = \sum_{\cv x} Q^{\otimes N}(\cv y^*|\cv x) p(\cv x) \geq \sum_{\cv x} \epsilon^N p(\cv x) = \epsilon^N$, we have
  \begin{IEEEeqnarray}{rCl}
    U(\cv y^*) 
    & = & - p_0 \sum_{\cv x} p(\cv x) \log (Q^{\otimes N}(\cv y^*|\cv x)-p_0)\nonumber \\
    & & + P(\cv y^*) \log \frac{P(\cv y^*)}{P(\cv y^*)-p_0} 
     + p_0 \log (P(\cv y^*)-p_0) \nonumber\\
    & & + \sum_{\cv x} p(\cv x) Q^{\otimes N}(\cv y^*|\cv x) \log \frac{Q^{\otimes N}(\cv y^*|\cv x)-p_0}{Q^{\otimes N}(\cv y^*|\cv x)} \\
    & \leq & - p_0  \log (\epsilon^N - p_0) + q^* \log \frac{\epsilon^N}{\epsilon^N-p_0}  \nonumber \\
    & & + p_0 \log (q^*-p_0) + q^* \log \frac{q^*-p_0}{q^*} \\
    & = & (q^*+p_0)\log \frac{q^*-p_0}{\epsilon^N-p_0} +  q^* \log\frac{ \epsilon^N}{q^*} \label{eq:up2}
  \end{IEEEeqnarray}
  The proof is completed by combining \eqref{eq:up1} and \eqref{eq:up2}.
\end{IEEEproof}

\begin{IEEEproof}[Proof of Lemma~\ref{lemma:1}]
  We relax $N$ to a real number and solve $\frac{\diff F(N)}{\diff N} = 0$, i.e.,
\begin{equation}\label{eq:4:8}
  1 - \epsilon^{N} + L N \epsilon^{N} \ln \epsilon = 0,
\end{equation}
  or
  \begin{equation}
    \epsilon^{-N} - 1 + L N \ln \epsilon = 0.
  \end{equation}
  Let $t = - N \ln \epsilon$, and denote by $t^*(L)$ the solution of $g(t) \triangleq e^t - 1 - L t  = 0, t>0$. Then the solution of \eqref{eq:4:8} is $N^* = t^*(L)/\ln (1/\epsilon)$.

  We know that $g(t) < 0$ for $0<t<t^*(L)$; and $g(t)>0$ for
  $t>t^*(L)$. Since $g(\ln L) = L-1-L\ln L <0$ and
  $g(2\ln L) = L^2 - 1 -2L\ln L >0$ when $L>1$, we have
  $\ln L < t^*(L) < 2 \ln L$ when $L > 1$. 
Last, using $\epsilon^{N^*} = e^{-t^*(L)}$, 
\begin{equation}
  0.25 \leq  \left(1-1/L\right)^L \leq (1-\epsilon^{N^*})^L \leq \left(1-1/{L^2}\right)^L < 1,
\end{equation}
and hence  $F(N^*) = \frac{(1-\epsilon^{N^*})^L}{N^*} = \frac{\ln\frac{1}{\epsilon} (1-\epsilon^{N^*})^L}{t^*(L)} = \Theta(\frac{\ln\frac{1}{\epsilon}}{\ln L})$.
\end{IEEEproof}

\begin{IEEEproof}[Proof of Lemma~\ref{lemma:z}]
  We group the elements of $\Sin$ into $\lceil |\Sin|/2 \rceil$ pairs, denoted collectively as $\Sin^{(2)}$, where each element of $\Sin$ appears in exactly one pair. When $|\Sin|$ is even, all pairs have distinct entries. When $|\Sin|$ is odd, exactly one pair has the two entries same and the other pairs have distinct entries. 
  
  For each pair $(x,x')\in \Sin^{(2)}$, fix $y_{x,x'}$ such that  $Q(y_{x,x'}|x)\geq \varepsilon_{Q}$ and $Q(y_{x,x'}|x') \geq \varepsilon_{Q}$.  Define $\mc Z$ as the collection of $z=(z_x,x\in \Qin)$ such that $z_{x}=y_{x,x'}$ and $z_{x'}=y_{x,x'}$ for all pairs $(x,x')\in \Sin^{(2)}$.
  Let $\Sout = \{y_{x,x'}:(x,x')\in \Sin^{(2)}\}$. Therefore, $|\Sout|\le \lceil |\Sin|/2 \rceil$. Hence
for any $x\in \Sin$ and $z\in \mc Z$, $\alpha(x,z)=z_x \in \Sout$. 
When $\mc A$ is even, 
\begin{IEEEeqnarray}{rCl}
  P(Z \in \mc Z) & = & \prod_{(x,x')\in \Sin^{(2)}} P(Z[x] = y_{x,x'}) P(Z_{x'}= y_{x,x'}) \\
  & = & \prod_{(x,x')\in \Sin^{(2)}} Q(y_{x,x'}|x) Q(y_{x,x'}|x')  \geq \prod_{(x,x')\in \Sin^{(2)}}   \varepsilon_Q^{2}=\varepsilon_Q^{|\Sin|}.
\end{IEEEeqnarray}
When $\mc A$ is odd,
\begin{IEEEeqnarray}{rCl}
  P(Z \in \mc Z) & = & \prod_{(x,x')\in \Sin^{(2)}:x\neq x'} P(Z[x] = y_{x,x'}) P(Z_{x'}= y_{x,x'}) \prod_{(x,x)\in \Sin^{(2)}} P(Z[x] = y_{x,x})  \\
  & = & \prod_{(x,x')\in \Sin^{(2)}:x\neq x'} Q(y_{x,x'}|x) Q(y_{x,x'}|x') \prod_{(x,x)\in \Sin^{(2)}} Q(y_{x,x}|x)
  \geq \varepsilon_Q^{|\Sin|}.%
\end{IEEEeqnarray}
\end{IEEEproof}

\section{Proofs about Achievability}
\label{app:proofs:ach}

\begin{IEEEproof}[Proof of Lemma~\ref{proposition:decoding:error:M1}]
  Suppose that the node $\ell-1$ transmits $u_\ell(x)$ for $N$ times, where $x\in\mc A$. We know that the entries of $\cv y_{\ell}$ are i.i.d. random variables with distribution $Q_{\ell}(\cdot\mid u_\ell(x))$.
The error probability for ML decoding at the node $\ell$ satisfies
\begin{IEEEeqnarray}{rCl}
\epsilon_{\ell}(x) & \leq &
P\left( \lor_{\overline{x}\ne x} 
\mathcal{L}_{\ell}(\overline{x}; \cv y_{\ell})
\ge
\mathcal{L}_{\ell}(x; \cv y_{\ell})
\right\}\\
&\le & \sum_{\overline{x}\in\mc A:~\overline{x}\ne x}
P\left(
\mathcal{L}_{\ell}(\overline{x}; \cv y_{\ell})
\ge
\mathcal{L}_{\ell}(x; \cv y_{\ell})
\right),
\end{IEEEeqnarray}
where the second inequality follows from the union bound.
For fixed $\overline{x}\in\mc A$ so that $\overline{x}\ne x$, we bound the probability $P\left(
\mathcal{L}_{\ell}(\overline{x}; \cv Y_{\ell})
\ge
\mathcal{L}_{\ell}(x; \cv Y_{\ell}) \right)$ by considering two cases.

If there exists a non-empty subset $\mc Y_0\subseteq\Qout$ so that for any $y_0\in\mc Y_0$, $Q_{\ell}(y_0\mid u_\ell(x))>0$ but $Q_{\ell}(y_0\mid u_\ell(\overline{x}))=0$, 
as long as $\cv y_{\ell}[i]\in\mathcal{Y}_0$ for some $i$, we can assert that $\mathcal{L}_{\ell}(\overline{x}; \cv y_{\ell})
<
\mathcal{L}_{\ell}(x; \cv y_{\ell})$.
Therefore, 
\begin{IEEEeqnarray}{rCl}
&&P\left(
\mathcal{L}_{\ell}(\overline{x}; \cv y_{\ell})
\ge
\mathcal{L}_{\ell}(x; \cv y_{\ell})
\right)
\le P\left(
\cv Y_{\ell}[i]\notin \mathcal{Y}_0, i=1,\ldots,N
\right) \\
& = & \left[\sum_{y\notin\mathcal{Y}_0}Q_{\ell}(y\mid u_\ell(x))\right]^{N}
=
\exp\left(
-N\log\frac{1}{\sum_{y\notin\mathcal{Y}_0}Q_{\ell}(y\mid u_\ell(x))}
\right),
\end{IEEEeqnarray}
where $\sum_{y\notin\mathcal{Y}_0}Q_{\ell}(y\mid u_\ell(x)) = 1 - \sum_{ y\in\mathcal{Y}_0}Q_{\ell}(y\mid u_\ell(x)) < 1$.

Otherwise, consider that the support of $Q_{\ell}(\cdot\mid u_\ell(x))$ belongs to the support of $Q_{\ell}(\cdot\mid u_\ell(\overline{x}))$. For $i=1,\ldots,N$, define the random variable $D_i = \log\frac{Q_{\ell}(\cv Y_{\ell}[i]\mid u_\ell(\overline{x}))}{Q_{\ell}(\cv Y_{\ell}[i]\mid u_\ell(x))}$. We see that $D_i$ are i.i.d., and satisfy
\begin{equation}
\log\varrho_{\ell}\le D_i\le -\log\varrho_{\ell},
\end{equation}
where $\varrho_{\ell}=\min_{x\in\Qin, y\in\Qout:Q_{\ell}(y\mid x)>0}Q_{\ell}(y|x)$, and
\begin{equation}
\mathbb{E}[D_i]= E_{\ell}' \triangleq  -\mathcal{D}_{\text{KL}}\left(
Q_{\ell}(\cdot\mid u_\ell({x}))\|Q_{\ell}(\cdot\mid u_\ell(\overline{x}))
\right),
\end{equation}
where $\mathcal{D}_{\text{KL}}$ denotes the Kullback-Leibler divergence.
We see that $E_{\ell}'>-\infty$. Moreover, as $u_\ell(x) \neq u_{\ell}(\bar x) \in \Qin^\ell$, $Q_{\ell}(\cdot\mid u_\ell({x}))\neq Q_{\ell}(\cdot\mid u_\ell(\overline{x}))$ and hence $E_{\ell}'\neq 0$. 
Applying Hoeffding's inequality, we obtain
\begin{IEEEeqnarray}{rCl}
  P\left(
\mathcal{L}_{\ell}(\overline{x}; \cv y_{\ell})
\ge
\mathcal{L}_{\ell}(x; \cv y_{\ell})
\right) & = & P\left(
\sum_{i=1}^{N}D_i\geq 0 \right)\\
& = &
P\left(
\sum_{i=1}^{N}\big(D_i-E_{\ell}'\big)\ge -NE_{\ell}'
\right)\\
&\le&
\exp\left(
-\frac{N E_{\ell}'^2 }{2\log^2\varrho_{\ell}}
\right).
\end{IEEEeqnarray}
The proof is completed by combining both cases. 
\end{IEEEproof}

\begin{IEEEproof}[Proof of Lemma~\ref{lm:r2}]
  Suppose $Q$ has size $m\times n$. As $C(Q)>\epsilon>0$, $m\geq 2$. Let $\cv a = (a_1,\ldots,a_n)$ be a row of $Q$, and construct a new $m\times n$ stochastic matrix $\tilde Q$ with all the rows $\cv a$. We have $C(\tilde Q) = 0$ and hence $|C(Q) - C(\tilde Q)| > \epsilon$. Since channel capacity as a function of stochastic matrices is uniformly continuous~\cite[Lemma I.1]{Niesen2007}, there exists a constant $\delta>0$ depending on $\epsilon$ such that $\|\tilde{Q} - Q\|_{\infty}>\delta.$
As a consequence, there exists another row $\cv a'=(a_1',\ldots,a_n')$ of $Q$ such that $\|\cv a -\cv a'\|_{\infty}>\delta$. Denote by $j$ the index such that $|a_j-a_j'|>\delta$.

Using the example of uniform reduction with $s=2$, we can choose $R$ so that $RQ$ is formed by $\cv a$ and $\cv a'$. Then we can find $W$ so that $RQW = U_2(\rho_1)$, where
\begin{equation}
\rho_1 = \sum_{k:a_{k}+a_{k}'>0} \frac{a_{k}^2}{a_{k}+a_{k}'} = 1 - \sum_{k:a_{k}+a_{k}'>0} \frac{a_{k}a_k'}{a_{k}+a_{k}'}.
\end{equation}
Based on the relation that
\begin{equation}
\frac{1}{2} - \sum_{k:a_{k}+a_{k}'>0} \frac{a_{k}a_k'}{a_{k}+a_{k}'}
=
\frac{1}{4}\sum_{k:a_{k}+a_{k}'>0} \frac{(a_{k}-a_k')^2}{a_{k}+a_{k}'}
\ge 
 \frac{1}{4}\frac{(a_{j}-a_j')^2}{a_{j}+a_{j}'}
 \ge \frac{\delta^2}{8},
 \end{equation}
we have the lower bound $\rho_1 \geq B$ with $B=\frac{1}{2} + \frac{\delta^2}{8}>1/2$.
For any $\varrho$ such that $1/2<\varrho\leq B$, we have $U_2(\varrho) = U_2(\rho_1)U_2(\frac{\rho_1+\varrho-1}{2\rho_1-1})$, and hence $RQWU_2(\frac{\rho_1+\varrho-1}{2\rho_1-1}) = U_2(\varrho)$.
\end{IEEEproof}

\begin{IEEEproof}[Proof of Lemma~\ref{lemma:red1}]
  As $\rank(Q) = r \geq s$, 
  we can find stochastic matrices $R$ and $W$ such that $\min\mathrm{inv}(RQW) = \kappa_s(Q)$. Let $B = (RQW)^{-1}$, and $K = B U_s(\varrho)$. As $RQWK = U_s(\varrho)$, we only need to show that for $1/s < \varrho \leq \rho_s(Q)$, $K$ is a stochastic matrix. Let $\cv 1$ be the all-one vector of certain length. We see that
$
    K \cv 1 = B U_s(\rho) \cv 1  = B \cv 1 = \cv 1,
$
  where the last equality follows because $RQW \cv 1 = \cv 1$ and $RQW$ is invertible.

  It remains to show that all the entries of $K$ are nonnegative.
  Let $b_{ij}$ be the $(i,j)$ entry of $B$. The $(i,j)$ entry of $K$ is
$k_{ij} = \frac{1}{s-1}\left[(1-\varrho) + b_{ij}(s\varrho-1)\right] \geq \frac{1}{s-1}\left[(1-\varrho) + \kappa_s(Q)(s\varrho-1)\right].$
  When $\kappa_s(Q)\geq 0 $, we have $k_{ij}\geq 0$ for any $\varrho \in (1/s, 1]$.
  When $\kappa_s(Q) < 0 $, we have $k_{ij}\geq 0$ for any $\varrho \in (1/s,\frac{\kappa_s(Q) -1}{s\kappa_s(Q) -1}]$.   
\end{IEEEproof}

\begin{IEEEproof}[Proof of Theorem~\ref{thm:bats}]
Recall the Markov chain relation in \eqref{eq:tran:rk}, where the transition matrix $\cv P$ is an $(M+1) \times (M+1)$ matrix with the $(i,j)$ entry ($0\leq i,j \leq M$):
\begin{equation}
  \label{eq:tm}
  p_{i,j} =
  \begin{cases}
    0 & i<j, \\
    \sum_{k=j}^N f(k;N,\epsilon)  \zeta_j^{i,k}  & i\geq j,
  \end{cases}
\end{equation}
where $f(k;N,\epsilon) = \binom{N}{k} (1-\epsilon)^k\epsilon^{N-k}$ is the probability mass function~(PMF) of the binomial distribution with parameters $N$ and $1-\epsilon$, and $\zeta_j^{i,k}$ is the probability that the $i\times k$ matrix with independent entries uniformly distributed over the field $\ff_q$ has rank $j$. We know that (ref. \cite[(2.4)]{yang17monograph}) $  \zeta_j^{i,k} = \frac{\zeta_j^i\zeta_j^k}{\zeta_j^jq^{(i-j)(k-j)}},$
where
\begin{equation}\label{eq:zeta}
  \zeta_r^m =
  \begin{cases}
    1 & r=0, \\
    (1-q^{-m})(1-q^{-m+1}) \cdots (1-q^{-m+r-1}) & 1\leq r \leq m.
  \end{cases}
\end{equation}
As shown in \cite{zhou19}, the matrix $\cv P$ admits the eigendecomposition $  \cv P = \cv V \cv \Lambda \cv V^{-1},$
where $\cv V = (v_{i,j})_{0\leq i,j \leq M}$ and $\cv \Lambda = \mathrm{diag}(\lambda_0,\lambda_1,\ldots,\lambda_M)$.
Here $\lambda_j=\sum_{k=j}^N f(k;N,\epsilon) \zeta_j^k$, $v_{i,j}=\zeta_j^i$ for $i\geq j$ and otherwise $v_{i,j}=0$.
It can be checked that $\lambda_0> \lambda_1 > \cdots > \lambda_M$.
Denote the $(i,j)$ entry $0\leq i,j \leq M$ of $V^{-1}$ by $u_{i,j}$.
We know that $u_{i,j}=0$ for $i<j$ and $u_{i,i} = 1/\zeta_i^i$.
Based on the formulation above, we have
\begin{IEEEeqnarray}{rCl}
  \E[\pi_L] & = & \pi_0 \cv V \cv \Lambda^L \cv V^{-1}
  \begin{bmatrix}
    0 & 1 & \cdots & M
  \end{bmatrix}^\top = \sum_{i=1}^M \lambda_i^Lv_{M,i} \sum_{j=1}^i j u_{i,j}  \\
  & = & \lambda_1^Lv_{M,1}u_{1,1} \left(1 + \sum_{i=2}^M \frac{\lambda_i^Lv_{M,i}}{\lambda_1^Lv_{M,1}u_{1,1}}\sum_{j=1}^i j u_{i,j}\right) \IEEEyesnumber \label{eq:app0} \\
  & = & \Theta(\lambda_1^L), \IEEEyesnumber \label{eq:app1}
\end{IEEEeqnarray}
where \eqref{eq:app0} follows from the fact that $v_{M,1}u_{1,1}>0$, and \eqref{eq:app1} is obtained by noting that
\begin{equation}
\sum_{i=2}^M \frac{\lambda_i^Lv_{M,i}}{\lambda_1^Lv_{M,1}u_{1,1}}\sum_{j=1}^i j u_{i,j}
=
o(1)
\end{equation}
as $\lambda_i\leq \lambda_1$ for $i\geq 2$.
By \eqref{eq:zeta}, we further have 
\begin{IEEEeqnarray}{rCl}
  \lambda_1 & = & \sum_{k=1}^N f(k;N,\epsilon) (1-q^{-k}) =  \sum_{k=1}^N f(k;N,\epsilon) - \sum_{k=1}^N f(k;N,\epsilon) q^{-k} \\
  & = & 1 -  f(0;N,\epsilon) -  \sum_{k=1}^N \binom{N}{k} (1-\epsilon)^k\epsilon^{N-k} q^{-k} = 1 - (\epsilon + (1-\epsilon)/q)^N.
\end{IEEEeqnarray}
The proof is completed.
\end{IEEEproof}

\end{document}